\documentclass[twocolumn]{aastex7}

\usepackage{mathtools}

\usepackage[multiple]{footmisc}
\usepackage{natbib}
\usepackage{CJK}

\received{}
\revised{}
\accepted{}
\shorttitle{Nearby Supernovae in the LSST Era}
\shortauthors{Sand et al.}

\graphicspath{{./}{figures/}}

\begin{document}

\title{Nearby Supernova Science in the Era of High Cadence, Deep Time Domain Surveys
}
 
\correspondingauthor{D. J. Sand}
\email{dsand@arizona.edu}

\newcommand{\TAMU}{\affiliation{Department of Physics and Astronomy, Texas A\&M University, 4242 TAMU, College Station, TX 77843, USA}}
\newcommand{\Mitchell}{\affiliation{George P.\ and Cynthia Woods Mitchell Institute for Fundamental Physics \& Astronomy, College Station, TX 77843, USA}}
\newcommand{\GSFC}{\affiliation{Astrophysics Science Division, NASA Goddard Space Flight Center, Mail Code 661, Greenbelt, MD 20771, USA}}
\newcommand{\UMD}{\affiliation{Joint Space-Science Institute, University of Maryland, College Park, MD 20742, USA}}
\newcommand{\UCB}{\affiliation{Department of Astronomy, University of California, Berkeley, CA 94720-3411, USA}}
\newcommand{\UCSB}{\affiliation{Department of Physics, University of California, Santa Barbara, CA 93106-9530, USA}}
\newcommand{\Monash}{\affiliation{School of Physics and Astronomy, Monash University, Clayton, Victoria 3800, Australia}}
\newcommand{\OzGrav}{\affiliation{OzGrav: The ARC Centre of Excellence for Gravitational Wave Discovery, Clayton, Victoria 3800, Australia}}
\newcommand{\noir}{\affiliation{NSF NOIRLab, 950 N Cherry Ave, Tucson, AZ 85719, USA}}
\newcommand{\noirHI}{\affiliation{NSF NOIRLab, 670 N. A’ohoku Place, Hilo, Hawai’i, 96720, USA}}
\newcommand{\noirctio}{\affiliation{Cerro Tololo Inter-American Observatory/NSF’s NOIRLab, Casilla 603, La Serena, Chile}}
\newcommand{\catalyst}{\altaffiliation{LSST-DA Catalyst Fellow}}
\newcommand{\UF}{\affiliation{Department of Astronomy, University of Florida, Bryant Space Science Center, Gainesville, FL
32611-2055, USA}}
\newcommand{\SETI}{\affiliation{SETI Institute, 339 Bernardo Ave, Suite 200, Mountain View, CA 94043, USA}}
\newcommand{\hunren}{\affiliation{HUN-REN CSFK, Konkoly Observatory, MTA Centre of Excellence, Konkoly Thege Miklós út 15-17, Budapest, 1121, Hungary}}
\newcommand{\TAU}{\affiliation{School of Physics and Astronomy, Tel Aviv University, Tel Aviv 69978, Israel}}
\newcommand{\ljmu}{\affiliation{Astrophysics Research Institute, Liverpool John Moores University, IC2, 146 Brownlow Hill, Liverpool, L3 5RF, UK}}
\newcommand{\UVA}{\affiliation{Department of Astronomy, University of Virginia, Charlottesville, VA22903, USA}}
\newcommand{\CMU}{\affiliation{McWilliams Center for Cosmology and Astrophysics, Department of Physics, Carnegie Mellon University, 5000 Forbes Avenue, Pittsburgh, PA 15213, USA}}
\newcommand{\IPAC}{\affiliation{IPAC, Mail Code 100-22, Caltech, 1200 E.\ California Blvd., Pasadena, CA 91125}}
\newcommand{\keck}{\affiliation{W.~M.~Keck Observatory, 65-1120 M\=amalahoa Highway, Kamuela, HI 96y43-8431, USA}}
\newcommand{\UCSD}{\affiliation{Department of Astronomy \& Astrophysics, University of California, San Diego, 9500 Gilman Drive, MC 0424, La Jolla, CA 92093-0424, USA}}
\newcommand{\CfA}{\affiliation{Center for Astrophysics \textbar{} Harvard \& Smithsonian, 60 Garden Street, Cambridge, MA 02138-1516, USA}}
\newcommand{\NU}{\affiliation{Department of Physics and Astronomy, Northwestern University, 2145 Sheridan Rd, Evanston, IL 60208, USA}}
\newcommand{\CIERA}{\affiliation{Center for Interdisciplinary Exploration and Research in Astrophysics (CIERA), \\Northwestern University, 1800 Sherman Ave., 8th Floor, Evanston, IL 60201, USA}}
\newcommand{\Thacher}{\affiliation{Thacher Observatory, Thacher School, 5025 Thacher Rd. Ojai, CA 93023, USA}}
\newcommand{\UA}{\affiliation{Steward Observatory, University of Arizona, 933 North Cherry Avenue, Tucson, AZ 85721-0065, USA}}
\newcommand{\UCDavis}{\affiliation{Department of Physics, University of California, 1 Shields Avenue, Davis, CA 95616-5270, USA}}
\newcommand{\Padova}{\affiliation{Department of Physics and Astronomy Galileo Galilei, University of Padova, Vicolo dell'Osservatorio, 3, I-35122 Padova, Italy}}
\newcommand{\INAF}{\affiliation{INAF Osservatorio Astronomico di Padova, Vicolo dell'Osservatorio 5, I-35122 Padova, Italy}}
\newcommand{\INAFbol}{\affiliation{INAF - Osservatorio di Astrofisica e Scienza dello Spazio - Via Piero Gobetti 93/3, I-40129 Bologna, Italy}}
\newcommand{\LPL}{\affiliation{Lunar and Planetary Lab, Department of Planetary Sciences, University of Arizona, Tucson, AZ 85721, USA}}
\newcommand{\NOAO}{\affiliation{National Optical Astronomy Observatory, 950 North Cherry Avenue, Tucson, AZ 85719, USA}}
\newcommand{\PATAU}{\affiliation{The School of Physics and Astronomy, Tel Aviv University, Tel Aviv 69978, Israel}}
\newcommand{\TTU}{\affiliation{Department of Physics and Astronomy, Texas Tech University, Box 1051, Lubbock, TX 79409-1051, USA}}
\newcommand{\OSU}{\affiliation{Department  of  Astronomy,  The  Ohio  State University,  140  W.  18th  Ave.,  Columbus,  OH43210, USA}}
\newcommand{\LBT}{\affiliation{Large Binocular Telescope Observatory, 933 North Cherry Avenue, Tucson, AZ, USA}}
\newcommand{\RMC}{\affiliation{Department of Physics and Space Science Royal Military College of Canada P.O. Box 17000, Station Forces Kingston, ON K7K 7B4, Canada}}
\newcommand{\ASU}{\affiliation{School of Earth and Space Exploration, Arizona State University, Tempe, AZ 85287, USA}}
\newcommand{\MMT}{\affiliation{MMT Observatory, PO Box 210065, University of Arizona, Tucson, AZ 85721-0065, USA}}
\newcommand{\NAU}{\affiliation{Department of Physics and Astronomy, Northern Arizona University, P.O. Box 6010, Flagstaff, AZ 86011, USA}}
\newcommand{\UAOptSci}{\affiliation{College of Optical Sciences, University of Arizona, 1630 E University Blvd, Tucson, AZ 85719, USA}}
\newcommand{\UNC}{\affiliation{Department of Physics and Astronomy, University of North Carolina at Chapel Hill, Chapel Hill, NC 27599, USA}}
\newcommand{\MSU}{\affiliation{Center for Data Intensive and Time Domain Astronomy, Department  of  Physics  and  Astronomy,  Michigan  State  University,East Lansing, MI 48824, USA}}
\newcommand{\UCSC}{\affiliation{Department of Astronomy and Astrophysics, University of California, Santa Cruz, CA 95064, USA}}
\newcommand{\STScI}{\affiliation{Space Telescope Science Institute, 3700 San Martin Drive, Baltimore, MD 21218, USA}}

\newcommand{\Rutgers}{\affiliation{Department of Physics and Astronomy, Rutgers, the State University of New Jersey,\\136 Frelinghuysen Road, Piscataway, NJ 08854-8019, USA}}

\newcommand{\Brandeis}{\affiliation{Department of Physics, Brandeis University, Waltham, MA 02453, USA}}
\newcommand{\LCO}{\affiliation{Las Cumbres Observatory, 6740 Cortona Drive, Suite 102, Goleta, CA 93117-5575, USA}}
\newcommand{\UToronto}{\affiliation{Department of Astronomy and Astrophysics, University of Toronto, 50 St. George Street, Toronto, Ontario, M5S 3H4 Canada}}
\newcommand{\NotreDame}{\affiliation{Department of Physics, University of Notre Dame, Notre Dame, IN 46556, USA}}
\newcommand{\UMN}{\affiliation{College of Science \& Engineering, Minnesota Institute for Astrophysics, University of Minnesota, 115 Union St. SE, Minneapolis, MN 55455, USA}}
\newcommand{\UT}{\affiliation{Department of Astronomy, University of Texas at Austin, Austin, TX 78712, USA}}
\newcommand{\JHU}{\affiliation{The Johns Hopkins University, Baltimore, MD 21218, USA}}
\newcommand{\VAT}{\affiliation{Vatican Observatory, 00120 Citt\`{a} del Vaticano, Vatican City State  }}
\newcommand{\HF}{\affiliation{Hubble Fellow}}
\newcommand{\Carnegie}{\affiliation{The Observatories of the Carnegie Institution for Science, 813 Santa Barbara St., Pasadena, CA 91101, USA}}

\author[0000-0003-4102-380X, gname=David, sname=Sand]{David J. Sand}
\email{dsand@arizona.edu}
\UA

\author[0000-0001-8073-8731, gname=Bhagya, sname=Subrayan]{Bhagya Subrayan}
\email{bsubrayan@arizona.edu}
\UA

\author[orcid=0000-0003-4175-4960]{Conor~L.~Ransome}
\UA
\email{cransome@arizona.edu}

\author[0000-0002-0832-2974, gname=Griffin, sname=Hosseinzadeh]{Griffin Hosseinzadeh}
\email{ghosseinzadeh@ucsd.edu}
\UCSD

\author[0000-0002-4924-444X]{K.\ Azalee Bostroem}
\IPAC
\email{bostroem@ipac.caltech.edu}

\author[0000-0003-0123-0062, gname=Jennifer, sname=Andrews]{Jennifer E. Andrews}
\email{jennifer.andrews@noirlab.edu}
\affiliation{Gemini Observatory/NSF's NOIRLab, 670 N. A'ohoku Place, Hilo, HI 96720, USA}

\author[orcid=0000-0002-0744-0047, gname=Jeniveve, sname=Pearson]{Jeniveve Pearson}
\email{jenivevepearson@arizona.edu}
\UCSD

\author[0000-0002-7937-6371]{Yize Dong %\begin{CJK*}{UTF8}{gbsn}(董一泽)\end{CJK*}
}
\CfA
\email{yize.dong@cfa.harvard.edu}

\author[0000-0001-8818-0795]{Stefano Valenti}
\UCDavis
\email{stfn.valenti@gmail.com}

\author[0000-0001-8738-6011]{Saurabh W.~Jha}
\email{saurabh@physics.rutgers.edu}
\Rutgers

\author[0000-0002-9454-1742, gname=Brian, sname=Hsu]{Brian Hsu}
\email{bhsu@arizona.edu}
\UA

\author[0000-0002-7352-7845]{Aravind P.\ Ravi}
\UA
\email{apazhayathravi@ucdavis.edu}

\author[0000-0002-4022-1874]{Manisha Shrestha}
\Monash\OzGrav
\email{Manisha.Shrestha@monash.edu}

\author[0000-0002-5740-7747, gname=Charles, sname=Kilpatrick]{Charles~D.~Kilpatrick}
\email{ckilpatrick@northwestern.edu}
\NU
\CIERA

\author[orcid=0000-0003-4537-3575, gname=Noah, sname=Franz]{Noah Franz}
\UA
\email{nfranz@arizona.edu}

\author[0000-0001-8341-3940]{Mojgan Aghakhanloo}
\affiliation{Department of Astronomy, University of Virginia, 530 McCormick Road, Charlottesville, VA 22904, USA}
\affiliation{Virginia Institute of Theoretical Astronomy, University of Virginia, Charlottesville, VA 22904, USA}
\email{}

\author[0000-0002-8977-1498]{Igor Andreoni}
\UNC
\email{igor.andreoni@unc.edu}

\author[0000-0002-1895-6639]{Moira Andrews}
\LCO\UCSB
\email{mandrews@lco.global}

%\author[0000-0001-7090-4898]{Iair Arcavi}
%\TAU
%\email{arcavi@tauex.tau.ac.il}
%\author[0009-0004-7268-7283]{Raphael Baer-Way}
%\affiliation{Department of Astronomy, University of Virginia, 530 McCormick Road, Charlottesville, VA 22904, USA}
%\affiliation{Department of Astronomy, University of Virginia,
% Charlottesville VA 22904-4325, USA}
%\email{placeholder@gmail.com}

\author[0000-0002-4449-9152]{Katie~Auchettl}
\affiliation{School of Physics, The University of Melbourne, Parkville, VIC 3010, Australia}
\email{}

%\author[0000-0003-4666-4606]{Emma R. Beasor}
%\ljmu
%\email{e.r.beasor@ljmu.ac.uk}

%\author[0000-0002-9392-9681, gname=Edo, sname=Berger]{Edo Berger}
%\email{eberger@cfa.harvard.edu}
%\CfA

\author[orcid=0000-0003-1953-8727]{Federica B. Bianco}
\affiliation{Department of Physics and Astronomy, University of Delaware, Newark, DE 19716, USA}
\affiliation{Joseph R. Biden, Jr. School of Public Policy and Administration, University of Delaware, DE 19716, USA}
\affiliation{University of Delaware, Data Science Institute, Newark, DE 19716, USA}
\affiliation{Vera C. Rubin Observatory, Tucson, AZ 85719, USA}
\email{}

%\author[0000-0003-0526-2248]{Peter Blanchard}
%\CfA
%\email{pblanchard@cfa.harvard.edu}

%\author[0000-0001-5955-2502]{Thomas G. Brink}
%\UCB
%\email{tgbrink@berkeley.edu}

%\author[0000-0002-1270-7666, gname=Tom\'as, sname=Cabrera]{Tom\'as Cabrera}
%\email{tcabrera@andrew.cmu.edu}
%\CMU

\author[orcid=0000-0003-4553-4033,gname=R\'egis, sname=Cartier]{R\'egis Cartier} 
\affiliation{Centro de Astronom\'ia (CITEVA), Universidad de Antofagasta, Avenida Angamos 601, Antofagasta, Chile}
\email{rgcartier@gmail.com}

\author[orcid=0000-0001-5078-5457,gname=Siddharth, sname=Chaini]{Siddharth Chaini} 
\altaffiliation{NASA FINESST Fellow}
\affiliation{Department of Physics and Astronomy, University of Delaware, Newark, DE 19716, USA}
\affiliation{University of Delaware, Data Science Institute, Newark, DE 19716, USA}
\email{chaini@udel.edu}

\author[0009-0007-7590-6914]{Diva Chaudhary}
\Monash\OzGrav
\email{dcha0095@student.monash.edu}

\author[orcid=0009-0007-2396-0003]{Analía V. Smith Castelli}
\affiliation{Instituto de Astrofísica de La Plata (CONICET - UNLP), Paseo del Bosque s/n, B1900FWA, La Plata, Argentina, Facultad de Ciencias Astronómicas y Geofísicas, Universidad Nacional de La Plata, Paseo del Bosque s/n, B1900FWA, La Plata, Argentina}
\email[]{}

%\author[0000-0002-7627-4839]{Adrian Crawford}
%\affiliation{Department of Astronomy, University of Virginia, 530 McCormick Road, Charlottesville, VA 22904, USA}
%\UVA
%\email{adrian.crawford@virginia.edu}

\author[orcid=0009-0006-4823-9768,gname=Shar, sname=Daniels]{Shar Daniels} 
\affiliation{Department of Physics and Astronomy, University of Delaware, Newark, DE 19716, USA}
\email{shard@udel.edu}

\author[0000-0003-3460-0103]{Alexei V. Filippenko}
\UCB
\email{afilippenko@berkeley.edu}

%\author[0000-0001-6395-6702]{Sebastian Gomez}
%\UT
%\email{sebastian.gomez@austin.utexas.edu}

\author[0000-0002-9154-3136]{Melissa L.\ Graham}
\affiliation{University of Washington, Dept. of Astronomy, Box 351580, Seattle, WA 98195, USA}
\affiliation{Institute for Data-intensive Research in Astrophysics and Cosmology, University of Washington, 3910 15th Avenue NE, Seattle, WA 98195, USA}
\email{mlg3k@uw.edu}

\author[0000-0002-9364-5419, gname=Xander, sname=Hall]{Xander J. Hall}
\email{xjh@andrew.cmu.edu}
\CMU

\author[0000-0002-1125-9187]{Daichi Hiramatsu}
\UF
\email{dhiramatsu@ufl.edu}

%\author[0009-0008-2052-8474]{\'Agoston Horti-D\'avid}
%\affiliation{ELTE E\"otv\"os Lor\'and University, Institute of Physics and Astronomy, 
%P\'azm\'any P\'eter s\'et\'any 1A, Budapest 1117, Hungary }
%\hunren
%\email{hortidavid.agoston@csfk.org}

\author[0000-0003-4253-656X]{D. Andrew Howell}
\affil{Las Cumbres Observatory, 6740 Cortona Drive, Suite 102, Goleta, CA 93117-5575, USA}
\affil{Department of Physics, University of California, Santa Barbara, CA 93106-9530, USA}
\email{dahowell@gmail.com}

%\author[0000-0001-7201-1938, gname=Lei, sname=Hu]{Lei Hu}
%\email{leihu@andrew.cmu.edu}
%\CMU

%\author[0000-0003-4131-5183]{Philip A. James}
%\ljmu
%\email{P.A.James@ljmu.ac.uk}

\author[0000-0003-3108-1328]{Lindsey~A.~Kwok}
\thanks{NHFP Hubble Fellow}
\CIERA
\email{lindsey.kwok@northwestern.edu}

%\author[0000-0001-5169-4143]{Gavin P. Lamb}
%\ljmu
%\email{G.P.Lamb@ljmu.ac.uk}

\author[0000-0002-6164-5051]{Amanda~R.~Lopes}
\affiliation{Instituto de Astronomia, Geofísica e Ciências Atmosféricas da Universidade de São Paulo, Cidade Universitária, CEP:05508-990, São Paulo, SP, Brazil}
\email[]{}

%\author[0000-0001-9589-3793, gname=Michael, sname=Lundquist]{Michael J. Lundquist}
%\email{mlundquist@keck.hawaii.edu}
%\keck

%\author[0000-0002-9144-7726]{Clara E. Martínez-Vázquez}
%\noirHI
%\email{ckilpatrick@northwestern.edu}

\author[0000-0001-6685-0479]{Thomas Matheson}
\noir
\email{tom.matheson@noirlab.edu}

\author[0000-0001-5807-7893]{Curtis McCully}
\LCO
\email{cmccully@lco.global}

\author[0000-0001-7132-0333]{Maryam Modjaz}
\affiliation{Department of Astronomy, University of Virginia, 530 McCormick Road, Charlottesville, VA 22904, USA}
%\affil{Department of Astronomy, University of Virginia, 530 McCormick Rd, Charlottesville, VA 22904, USA}
\email{placeholder@email.com}

\author[0009-0006-0647-636X]{Phillip Noel}
\UA
\email{phillipnoel@arizona.edu}

\author[orcid=0000-0002-6639-6533,gname='Gregory', sname='Paek']{Gregory S. H. Paek}  
\affiliation{Institute for Astronomy, University of Hawai`i, 2680 Woodlawn Drive, Honolulu, HI 96822, USA}
\email{gregorypaek94@gmail.com}

\author[0000-0002-6011-0530, gname=Antonella, sname=Palmese]{Antonella Palmese}
\email{apalmese@andrew.cmu.edu}
\CMU

%\author[0009-0000-5120-1193]{Avi Patel}
%\UCSC
%\email{avpapate@ucsc.edu}

\author[0000-0001-6806-0673, gname=Anthony, sname=Piro]{Anthony L.\ Piro}
\email{piro@carnegiescience.edu}
\Carnegie

%\author[0000-0001-8907-3051]{Joanne L.\ Pledger}
%\affiliation{Jeremiah Horrocks Institute, University of Lancashire, Preston, PR1 2HE, UK}
%\email{jpledger@lancashire.ac.uk}

%\author[0000-0003-3643-839X]{Jeonghee Rho}
%\SETI
%\email{jrho@seti.org}

%\author[0000-0003-0926-3950]{Kriszti\'an S\'arneczky}
%\hunren
%\email{sarneczky.krisztian@csfk.org}

\author[0000-0001-6360-992X]{Monika Soraisam}
\noirHI
\email{monika.soraisam@noirlab.edu}

%\author[0000-0002-1468-9668]{Jay Strader}
%\MSU
%\email{straderj@msu.edu}

%\author[0000-0001-5567-1301]{Francisco Valdes}
%\noir
%\email{frank.valdes@noirlab.edu}

\author[0000-0002-4951-8762]{Sergiy Vasylyev}
\UCSD
\email{svasylyev@ucsd.edu}

\author[0000-0001-8764-7832]{J\'ozsef Vink\'o}
%\affiliation{HUN-REN Research Centre for Astronomy and Earth Sciences, Konkoly Observatory, Konkoly Th. M. {\'u}t 15-17., 1121 Budapest, Hungary}
%\affiliation{ELTE E\"otv\"os Lor\'and University, Institute of Physics and Astronomy, P\'azm\'any P\'eter s\'et\'any 1, Budapest, Hungary}
\hunren
\affiliation{Department of Experimental Physics, Institute of Physics, University of Szeged, D{\'o}m t{\'e}r 9, 6720 Szeged, Hungary}
\email{vinko@konkoly.hu}

%\author[0000-0003-4341-6172]{A.~Katherina~Vivas}
%\noirctio
%\email{kathy.vivas@noirlab.edu}

%\author[0000-0002-5814-4061]{V.~Ashley~Villar}
%\affiliation{Center for Astrophysics \textbar{} Harvard \& Smithsonian, 60 Garden Street, Cambridge, MA 02138-1516, USA}
%\affiliation{The NSF AI Institute for Artificial Intelligence and Fundamental Interactions}
%\email{ashleyvillar@cfa.harvard.edu}

%\author[0000-0001-7092-9374]{Lifan Wang}
%\TAMU
%\email{lifan@tamu.edu}

\author[0000-0003-1349-6538]{J. Craig Wheeler}
\UT
\email{wheel@astro.as.utexas.edu}

\author[0000-0003-4537-3575]{Samuel D. Wyatt}
\GSFC
\email{samuel.d.wyatt@nasa.gov}

%\author[orcid=0009-0006-7296-728X]{Kathryn Wynn}
%\LCO \UCSB 
%\email{}

%\author[0000-0001-5955-2502]{WeiKang Zheng}
%\UCB
%\email{weikang@berkeley.edu}

\begin{abstract}
We discuss science opportunities for nearby supernovae (SNe) in the era of the Vera C. Rubin Observatory’s Legacy Survey of Space and Time (LSST), emphasizing the youngest phases of SN evolution and the value of contemporaneous high-cadence ``shadowing'' surveys.  LSST's Wide Fast Deep survey will discover enormous numbers of SNe, but its $\sim 3$ day cadence will miss signatures that evolve on timescales of hours to days, requiring cadence augmentation by other programs.  We present simple depth and cadence metrics for early SN signatures and use empirical nearby SN discovery rates to estimate the resulting science opportunities. LSST will detect or set limits on precursor emission from hundreds of core-collapse SNe, including $\gtrsim 50$--100 Type IIn {\it and} normal Type II SNe annually, as well as $\gtrsim$5 Type Ibn.   Within $\sim 200$ Mpc, roughly 250 Type Ia SNe are classified annually, providing the potential to identify hundreds of early light-curve excesses, and a robust measurement of the incidence  of extreme high-velocity features ($>$25,000 km s$^{-1}$). A shadowing survey with a depth of $\sim 22.5$ mag could search for early light-curve features, akin to the light-curve excess seen in SN\,2023ixf, across dozens of nearby core-collapse SNe every year. Rapid spectroscopy of $\gtrsim$150 Type II SNe per year within $\sim$160 Mpc could constrain the incidence of short-lived ``flash'' features to $\lesssim$10\% within a few years, while also tying this to the rate of precursor emission in normal Type II SNe.  We also discuss the challenges posed by early classification ambiguity and faint nonterminal transients, and we request that the 80 hour embargo on LSST images be waived for nearby galaxy fields. Finally, we advocate for a Nearby Galaxy Transient Broker that integrates transient streams, archival data, forced photometry, and follow-up triggering in a single platform.
\end{abstract}

%(possibly signaling a nondegenerate companion, circumstellar material, or an unusual nickel distribution)

\keywords{Supernovae (1668), Core-collapse supernovae (304), Type Ia supernovae (1728), Sky surveys (1464)}

\section{Introduction} \label{sec:intro}

We are in the ``golden age'' of explosive, time-domain astrophysics.  Multiple transient surveys now operate simultaneously over large fractions of the sky, releasing (many of) their candidates in nearly real time. Multimessenger astronomy promises new breakthroughs from the discovery and analysis of the counterparts to gravitational-wave and neutrino events.  As a capstone, the Vera C. Rubin Observatory's Legacy Survey of Space and Time (LSST; \citealt{lsst}) has begun operations and will report $\sim$10$^{6}$ transient and variable objects per night, surveying the accessible southern sky every $\sim$3 days in six filters via the Wide Fast Deep (WFD) survey.  Given the excitement and scientific promise of time-domain astrophysics, the Astro2020 Decadal Survey identified ``New Windows on the Dynamic Universe'' as one of three science priority areas \citep{Decadal}.
 
The deluge of time-domain data from LSST in particular will spur a wide variety of investigations, ranging from eruptive transients and variables within the Local Group of galaxies to huge collections of light curves of ordinary and rare supernovae (SNe) out to moderate redshift.  The time variability of active galactic nuclei, tidal disruption events, and other nuclear phenomena will be another focus.  Many of these science cases have been outlined in the Roadmap of the Transient and Variable Stars Science Collaboration \citep{TVS23}, while the use of Type Ia SNe for cosmology is discussed in many publications \citep[e.g.][]{desc_roadmap}.  LSST will also spend up to 3\% of its time devoted to discovering the counterparts to multimessenger events \citep{RubinToO}.
%Era of time domain astronomy. 2020 Decadal. LSST.

Despite the large increase in transient and supernova (SN) numbers in recent times, {\it the intensive study of individual SNe drives the understanding of their explosions and progenitor systems, and the nearest and brightest objects are the ones that can be studied most intensively}.  This is especially true when new types of observations are made that can stress theoretical models and unveil new physical phenomena (e.g., in the moments after explosion, at high cadence, in new wavelength regimes, etc.).  Some recent examples of this have included the early light-curve features and excesses seen in thermonuclear \citep[e.g.,][among others]{Marion16,Hosseinzadeh17,Hosseinzadeh22,Hosseinzadeh_23bee,Dimitriadis19_18oh,Shappee19,Burke21,Burke25,Pearson24,Wang24} and core-collapse SN light curves \citep{Tartaglia17,Armstrong21,Hosseinzadeh23,Das23,Shrestha24,Subrayan25}. New and/or underexplored features are also seen in spectra; examples include the extreme early velocities seen in some Type Ia SNe \citep{Hosseinzadeh22,Hosseinzadeh_23bee} and ``flash'' features observable in the first hours to days after a normal core-collapse explosion \citep[e.g.,][]{Galyam14,Smith15,Yaron17,khazov16,bruch21,bruch23,JG22,JG23,JG24,JG24_SN24ggi,Tartaglia21,Terreran22,Bostroem23,Shrestha24,Pessi24,Andrews25}.  Likewise, space-based ultraviolet \citep[UV; e.g.,][]{Bostroem23,Bostroem24,Zimmerman24}  and mid-infrared (MIR) spectroscopy \citep[e.g.][]{Kwok23,Kwok24,Kwok25,Kwok26a, Kwok26b,Derkacy23,Derkacy24,Derkacy25,Ashall24,Baron25,Medler25,Mera26, Macrie26, Hall26, Baron26,Shrestha26} provide unique insights into the SN explosion and its environment, but data with high signal-to-noise ratio (S/N) can only be obtained for the nearest and brightest objects.  In this paper, we argue that exceptional observations of nearby SNe will become more common, especially with early warning from multiple time-domain surveys. % and prompt, sustained data is collected.  

The LSST 
%Wide-Fast-Deep (WFD) 
WFD survey will have a cadence of $\sim$3 days and observe $\sim$18,000 deg$^2$ over $\sim$800 epochs during the 10\,yr survey \citep{lsst,Bianco22}.  Despite the depth of the WFD ($\sim$24-25 mag per epoch; see below) and the areas covered, the cadence of the survey will be the limiting factor for finding young SNe in the nearby universe; often SN signatures within a day of first light are key to constraining the progenitor and explosion mechanism.  To fill this gap, there is an opportunity to ``shadow'' LSST in the hours to day after a visit in order to identify the youngest transients.  This will inherently be done by ongoing time-domain surveys, and focused experiments may shadow LSST with instruments such as the Dark Energy Camera (DECam) on the 4\,m Victor M. Blanco telescope \citep[][]{decam}, a natural fit given its aperture and 3 deg$^2$ field of view.  Indeed, the Shadow Survey has begun such a program to combine LSST and DECam observations around nearby galaxy clusters in order to find nearby SNe within a day of explosion at unprecedented, faint luminosities \citep{Ransomeshadow}.

Here we use a handful of known core-collapse and thermonuclear SNe with exceptional data to showcase the observational signatures that will become more common in the LSST era. These events challenge existing models and have revealed new phenomena, providing a preview of the science enabled by combining and closely monitoring multiple survey streams. Discussing these opportunities, and the associated time-domain ecosystem, is the goal of the current work.  In Section~\ref{sec:landscape} we present the time-domain survey landscape and relevant infrastructure as critical context.  Section~\ref{sec:overview} discusses the broad categories of early SN observational signatures and connects them to survey depths and cadences, while in Sections~\ref{sec:ccsn} and \ref{sec:thermo} we respectively discuss core-collapse and thermonuclear SN signatures in more depth.  Challenges and possible solutions for acquiring and interpreting early SN data are outlined in Section~\ref{sec:discussion}, and we conclude in Section~\ref{sec:summary}.

 \begin{figure*}
\centering
\includegraphics[width=0.95\textwidth]{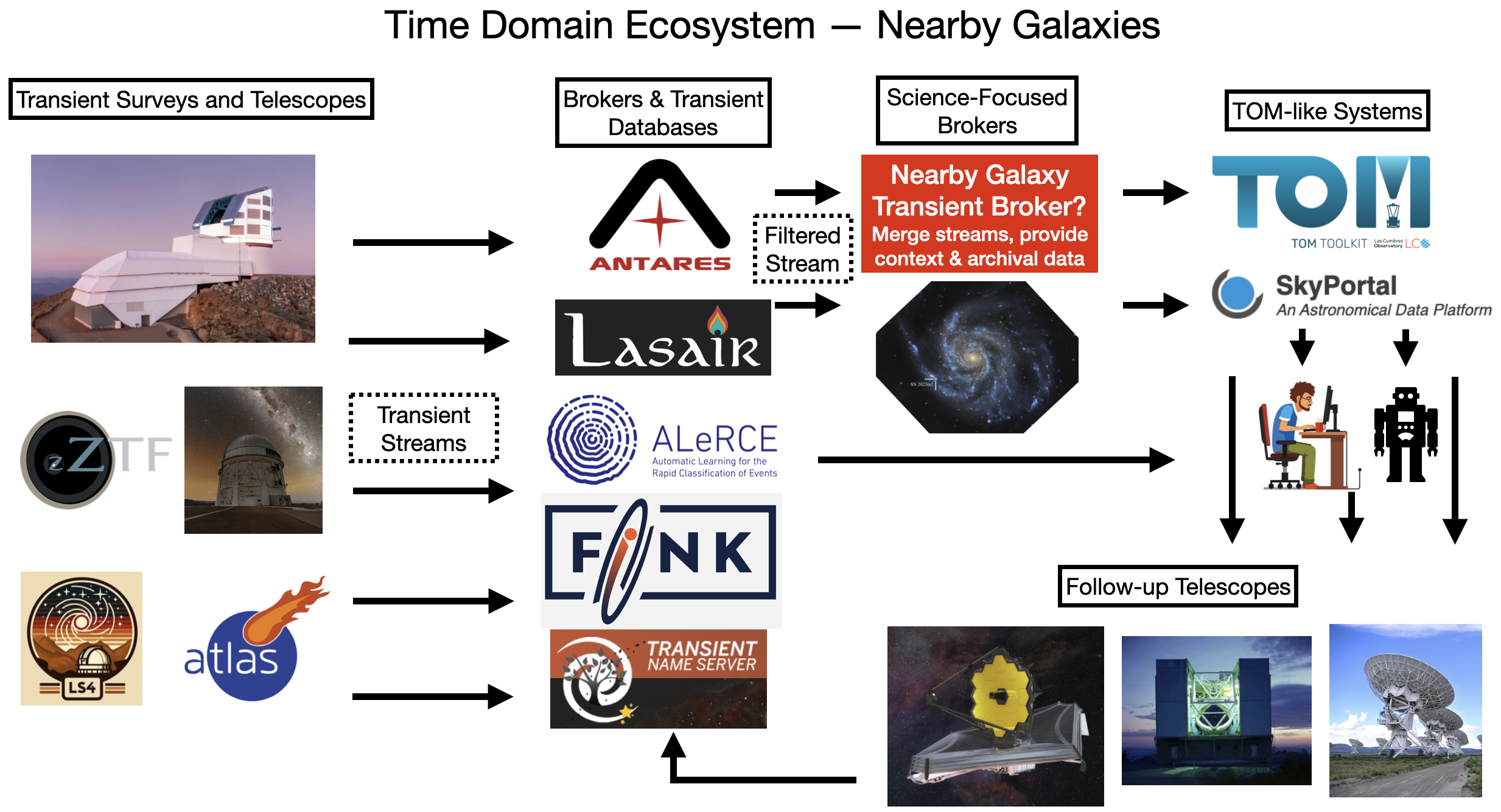}
\caption{Schematic of the time-domain ecosystem, where transients are discovered by various surveys {\it (left)} and streamed to brokers. Filtered sets of these transients are then sent to observers and their TOMs, or TOM-like systems.  A possible intermediate step involves filtered streams being sent to a science-focused broker; for instance, a ``Nearby Galaxies'' broker may only ingest the set of transients around galaxies within a distance limit (e.g., all galaxies within $<$200\,Mpc).  These science-focused brokers may merge transient streams or supply extra contextual or archival information relevant for a given science topic, as we discuss in Section~\ref{sec:discussion} for the nearby galaxy science case.  Once targets are ingested into a TOM, follow-up observations may be triggered either by a human or bot, which can then feed this new information back into the broker system. \label{fig:schematic}}
\end{figure*}

% One is the intensive study of the nearest and brightest transients.  This can be transformative in many ways... also amenable to space telescope follow-up.

\section{The Time-Domain Landscape}
\label{sec:landscape}
LSST will be the pre-eminent time-domain discovery engine for the next decade.  Other transient programs will also make important contributions, as they can augment the cadence of LSST (or cover the entire northern hemisphere, which LSST cannot), or observe in complementary wavebands.  Along with this suite of discovery engines, a whole ecosystem of time-domain infrastructure is being built to ingest transient alerts from the surveys, put them in context, conduct initial science analyses, and request follow-up observations from multiwavelength facilities.  We present a schematic of the time-domain ecosystem in Figure~\ref{fig:schematic}, giving an overview of the various components. Here we briefly summarize this landscape with an emphasis on how it may facilitate nearby SN discovery and rapid observations (along with areas where it may be inadequate).  We discuss the relevant observational signatures in subsequent sections.

\subsection{LSST}

The Vera C. Rubin Observatory is an 8-m telescope whose sole instrument is LSSTCam, a 9.6 deg$^2$ imager. %The Legacy Survey of Space and Time 
LSST will be conducted on the telescope over 10\,yr. The most relevant LSST program for this paper is the WFD survey, which will observe with a $\sim$3-day cadence in two filters per visit. The per-visit depth (5$\sigma$) of each image is expected to be ($u$,$g$,$r$,$i$,$z$,$y$)=(23.9,25.0,24.7,24.0,23.3,22.1) mag \citep{Bianco22}.  For our discussions below, which involve transients embedded in complicated nearby galaxy backgrounds, we conservatively assume a nominal transient detection depth of $g$,$r$ = 24.0 mag.

Key to the success of real-time transient studies with LSST is the rapid dissemination of new candidates to the community, performed by the Prompt Processing pipelines.  For more information on these data products, we refer the reader to \citet{Graham22,RTN-011,10.71929/rubin/3020138}, although relevant aspects are summarized here.  Transient detection and data products will be delivered as an alert stream to a set of seven full-stream alert brokers (see below) within 60\,s of image readout.  These data include image stamps (of the science region, difference and template images), flux measurements of the variable object, and flux measurements over the last year at that sky position (or noise measurements if no source was detected in the difference image during that time).  
The full processed visit and difference images that result from the Prompt Processing pipelines are subject to an 80 hour embargo required by the funding agencies, the National Science Foundation and the Department of Energy \citep{LDO-13,DMTN-199}. The embargo is an issue when close examination of the data is required to do very early-time science. %The actual images that result from the Prompt Processing pipelines will be available only after 80\,hr, which may be an issue when close examination of the data is required to do very early-time science. %, as we will discuss further below.

A ``forced-photometry service'' akin to that available from other time-domain surveys \citep[e.g., ATLAS;][]{Smith20,Shingles21} will allow detection limits to be determined at a given sky location where no alert has been issued, but these data will be subject to the 80\,hr restriction as well, complicating efforts to detect low-significance SN emission at the very earliest times.

\subsection{Other Optical/Near-Infrared Surveys}

Several wide-field transient surveys will be in operation alongside LSST, including the All-Sky Automated Survey for Supernovae \citep[ASAS-SN;][]{Shappee14}, the Panoramic Survey Telescope And Rapid Response System \citep[PanSTARRS][]{Kaiser10}, the La Silla Schmidt Southern Survey \citep[LS4;][]{LS4}, the Asteroid Terrestrial-impact Last Alert System \citep[ATLAS;][]{tonry18}, the Zwicky Transient Facility \citep[ZTF;][]{ztf}, the Young Supernova Experiment \citep[YSE;][]{Jones21}, BlackGEM \citep{Blackgem2}, and the Gravitational-Wave Optical Transient Observer \citep[GOTO;][]{Steeghs22}.  Impactful galaxy-targeted programs such as the Distance Less Than 40 Mpc survey \citep[DLT40;][]{Tartaglia18} and the amateur observer community will continue their efforts. In the medium term, powerful time-domain surveys will continue to come online, such as the Argus Optical Array \citep{argus}.  We also note that early in the LSST survey, transients may be detected by using archival DECam images as templates \citep[e.g., the {\sc SLIDE} software package;][]{Dong25}. These programs have an array of survey depths, area covered, cadences, and science goals --- we summarize their nominal cadence and depth in Table~\ref{tab:surveys}. It should be kept in mind that the values in Table~\ref{tab:surveys} can vary if science priorities for a given project change.  We also list a fiducial DECam Shadowing program, assuming deep imaging will be taken of select WFD fields $\lesssim$1 day after LSST has observed \citep[e.g.,][]{Ransomeshadow}.

If transients are reported in real time, these programs can be used in combination with each other and LSST to obtain extremely high-cadence observations of nearby galaxy fields for early discovery.  Indeed, this is an explicit science goal of several surveys (LS4, \citealt{LS4}; ZTF, \citealt{Kasliwal25}; YSE \citealt{Jones21}, although only LS4 is a southern sky survey). Despite this opportunity, it will be challenging to combine the data streams of multiple surveys in real time to find the youngest transients (see below).  A key metric is the relative depth of each survey, which translates to the absolute magnitude (or luminosity) a program reaches for a given distance, and thus the early observational signatures of the progenitor and explosion to which they are sensitive.  We list the absolute magnitude that each survey can reach for fiducial distances of $D$ = 20, 50, and 100\,Mpc in Table~\ref{tab:surveys}.  These values can be paired with the observational signatures we describe in Sections~\ref{sec:ccsn} and \ref{sec:thermo}.

\begin{deluxetable*}{lcccccc}\label{tab:surveys}
\tablecolumns{5}
\tablewidth{0pt}\tablecaption{Time-Domain Survey Cadence and Depth}
\tablehead{
\colhead{Survey Name} & \colhead{Typical Depth} & \colhead{Typical Cadence} & \multicolumn{3}{c}{Absolute mag at}\\
\colhead{} & \colhead{} & \colhead{{(days)}} & \colhead{20 Mpc} & \colhead{50 Mpc} & \colhead{100 Mpc}} 
\startdata
\hline
\hline
LSST WFD & $g,r,i$ $\approx$ 25.0,24.7,24.0\tablenotemark{a} & 3 & $-$7.5 & $-$9.5 & $-$11.0\\
%&$\approx$24.0 for nearby transients &&  \\
DECam Shadow & $g,r$ $\approx$ 22.5 & 1 & $-$9.0 & $-$11.0 & $-$12.5 \\
\hline
ASASSN & $g$ $\approx$ 18.0 & 1 & $-$13.5 & $-$15.5 & $-$17.0\\
ATLAS & $o,c$ $\approx$ 19.0,19.5 & 1 & $-$12.0 & $-$14.0 & $-$15.5\\
BlackGEM & $i,q$ $\approx$ 20.0,21.0\tablenotemark{a} & $\lesssim$1\tablenotemark{b} & $-$11.5 & $-$13.5 & $-$15.0 \\
DLT40 & $r$ $\approx$ 19.0 & 0.5--1 & $-$12.5 & $-$14.5 & $-$16.0 \\
GOTO & $L$ $\approx$ 19.5 & $\sim$3  & $-$12.0 & $-$14.0 & $-$15.5\\
LS4 & $g,i,z$ $\approx$ 20.5 & 3,1\tablenotemark{b} & $-$11.0 & $-$13.0 & $-$14.5\\
YSE & $g,r,i$ $\approx$ 21.5 & $\sim$3 & $-$10.0 & $-$12.0 & $-$13.5 \\
ZTF & $g,r,i$ $\approx$ 21.3,21.0,20.0\tablenotemark{a} & $\sim$1--3\tablenotemark{b} & $-$10.5 & $-$12.5 & $-$14.0\\
\hline
\enddata
\tablenotetext{a}{We conservatively assume $\sim$24 mag for transient detection limits from LSST throughout this work.  We also assume $c \approx 19.5$ mag for ATLAS, $i \approx 20.0$ mag for BlackGEM, and $r \approx 21.0$ for ZTF.}
\tablenotetext{b}{The short cadence is for the BlackGEM Local Transient Survey ($<$15 Mpc).  LS4 will have a 1\,day cadence search for fast optical transients and a 3\,day extragalactic SN cadence. The public ZTF survey has a 3\,day cadence, although they will survey the accessible LSST footprint with a 1\,day cadence \citep{Kasliwal25}.}
\end{deluxetable*}

\subsection{Time-Domain Infrastructure}

Given the proliferation of time-domain surveys, a whole ecosystem of tools has been built to stream transient candidates from the surveys and put them in astrophysical context (e.g., host galaxy, association with a known variable object).  This information can then be used to trigger observations based on preset filters or other means.  We show a schematic of this ecosystem in Figure~\ref{fig:schematic}.

The community alert brokers will ingest all incoming LSST transients, and may also ingest data from other transient surveys with real-time streams (this is limited at the present time).  These alert brokers are {\sc fink} \citep{fink}, AMPEL \citep[Alert Management, Photometry, and Evaluation of Light Curves;][]{ampel}, Babumul/BOOM \citep{boom}, ALeRCE \citep[Automatic Learning for the Rapid Classification of Events;][]{Alerce}, {\it Lasair} \citep{lasair}, the Pitt-Google Broker \citep{pitt}, and ANTARES \cite[Arizona-NOIRLab Temporal Analysis and Response to Events System;][]{antares}.  The Transient Name Server\footnote{\url{https://www.wis-tns.org/}} (TNS) is another vital piece of infrastructure, as it is the official IAU mechanism for reporting new astronomical transients; these may be promoted by the previously mentioned brokers, or come straight from a given transient survey.  Other venues for reporting new transients, or providing deeper information about them, include TNS-derived AstroNotes \citep{astronote}, GCN Notices and Circulars \citep{GCN}, and HERMES\footnote{\url{https://hermes.lco.global/}}, a web interface to Kafka\footnote{\url{https://kafka.apache.org}} streams for time domain and multi-messenger astronomy. In principle, the data stream from the brokers and TNS can be filtered to allow users to identify objects from a specific galaxy list, or based on the color or rise time of a given transient  (as a few examples).  However, most of the current transient brokers only have plans to ingest LSST and ZTF alerts, and to do so without necessarily ``merging'' these streams. There is a real need for a unified transient data stream and service (both ANTARES and Babumul do this for LSST and ZTF), which would simplify the identification of young transients in the nearby Universe. 

\begin{figure*}
\centering
\includegraphics[width=0.49\textwidth]{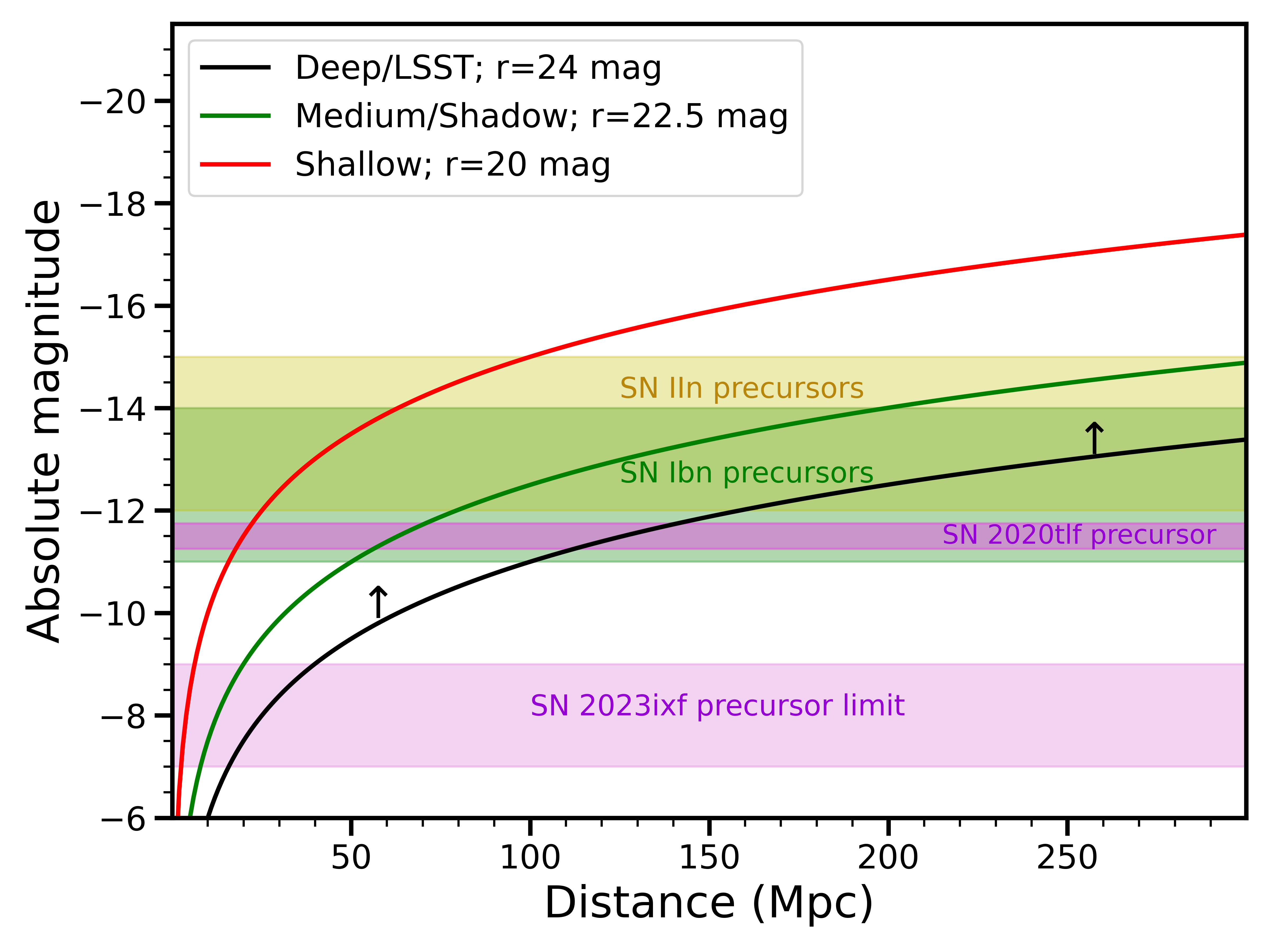}%\vspace{-20pt}
\includegraphics[width=0.49\textwidth]{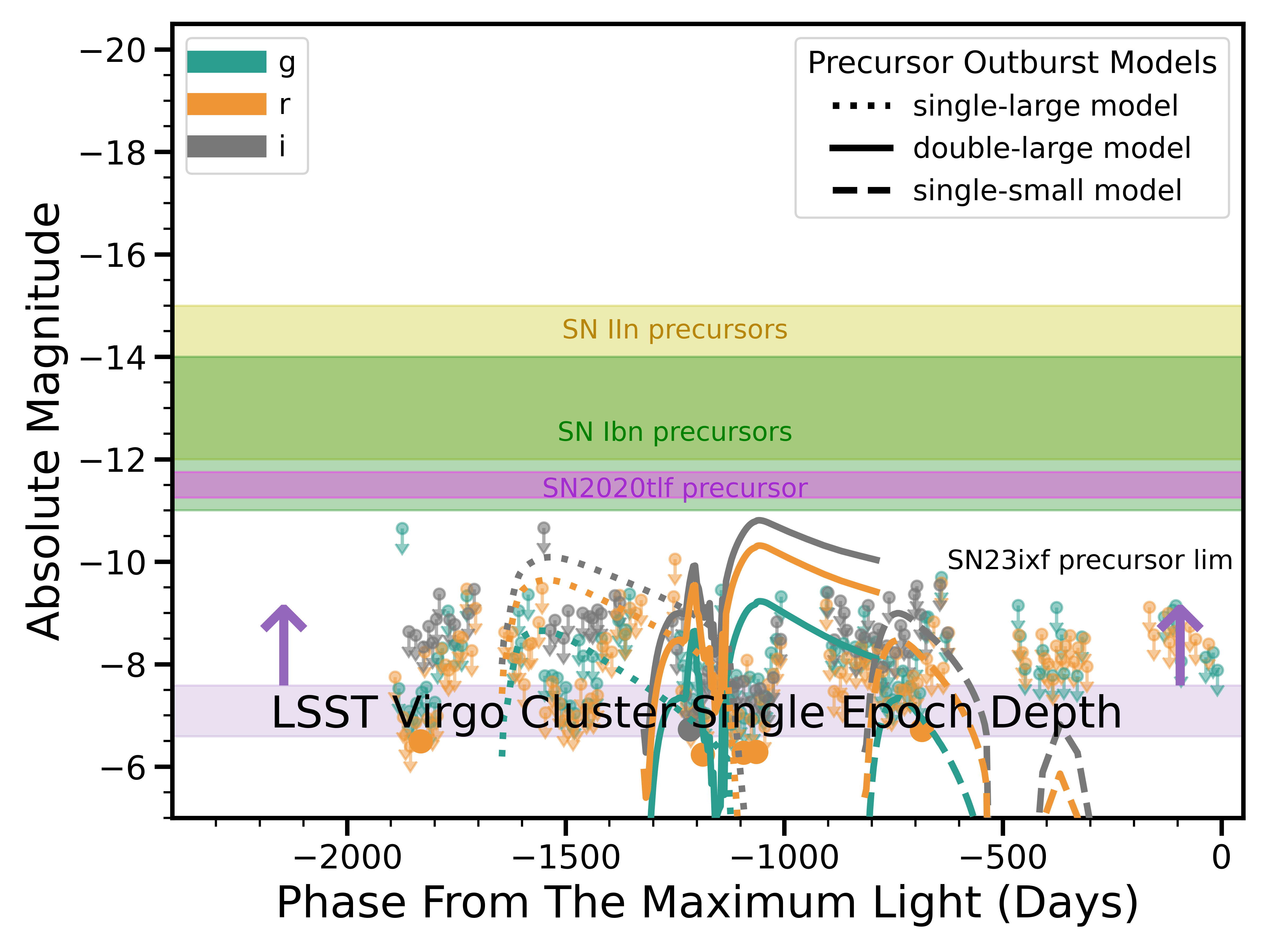}
\caption{Approximate survey sensitivities to SN precursor outbursts and brightenings. {\it Left:} Survey sensitivity to precursor emission and outbursts as a function of transient distance. See Tables~\ref{tab:surveys} and \ref{tab:SNsignatures} for fiducial values for survey depths and typical precursor luminosities. {\it Right:}  Precursor detection limits for SN\,2023ixf as a function of time prior to explosion \citep[see][for details]{Dong23}, alongside precursor outburst models from \citet{Tsuna23}. We also illustrate fiducial values for typical precursor luminosities.  Along the bottom, in light purple, we show the typical single-epoch depth of the LSST WFD survey at the distance of the Virgo Cluster, which will be sensitive to all of these possible signatures. \label{fig:precursor}}
\end{figure*}

The next layer in the time-domain software ecosystem are Target and Observation Managers (TOMs), which are sophisticated data-management platforms which ingest relevant or interesting portions of the broker data stream in order to request  follow-up observations and perform scientific analysis and visualization for targets of interest.  These systems may ingest additional catalogs and information to put transients into context, and may build tools to easily analyze spectra or light curves as new data are taken.  They may also have software interfaces that allow for the fast or automated triggering of new observations at a variety of multiwavelength facilities in response to incoming transient data --- a feature that is critical for acquiring early-time data on nearby transients. The open-source TOM Toolkit \citep{TOM} provides a backbone code base to customize a given TOM's infrastructure for different science-use cases, which has been adopted by the Las Cumbres Observatory's Supernova Exchange\footnote{\url{https://supernova.exchange/}}, SAGUARO \citep[Searches after Gravitational Waves Using Arizona's Observatories;][]{Hosseinzadeh24}, Black Hole TOM \citep{bhtom}, and GOATS \citep[Gemini Observation and Analysis of Targets System;][]{goats} teams.  Other purpose-built data platforms with TOM-like properties include the PESSTO marshal \citep{pessto}, {\sc SkyPortal} \citep{skyportal}, {\sc YSE-PZ} \citep{ysepz},  and Astro-COLIBRI \citep{colibri}. 

There are several automated and dedicated photometry and spectroscopy programs that tie into the alert broker and TOM ecosystem to obtain essential complementary data, often in real time. Some of these have already been operational, such as the Global Supernova Project (GSP) at Las Cumbres Observatory \citep{Brown13} and the Public European Southern Observatory Spectroscopic Survey of Transient Objects \citep{pessto}.  Others are ramping up as LSST begins, such as the SOXS collaboration \citep{SOXS}, TiDES \citep[the 4MOST Time Domain Extragalactic Survey;][]{tides}, and the Public AEON Spectroscopic Survey for Transient Astronomy \citep[PASSTA;][]{passta}.
Some newer programs are obtaining  follow-up observations in an automated fashion, closing the time between discovery and intensive monitoring even further, such as BTSBot \citep{btsbot1,btsbot} and the  DLT40 program \citep{Tartaglia18}, which also makes use of {\sc PyMMT}, an API that allows rapid spectroscopic requests at the 6.5\,m MMT telescope using the Binospec or MMIRS spectrographs \citep{Shrestha_MMT}.  In addition, older facilities like the ANU 2.3m telescope are being retrofitted for autonomous observations \citep{anu_automate}. The Neil Gehrels Swift Observatory also offered ``Priority 0'' observations for selected guest-investigator programs to trigger extremely fast-response UV light curves (within minutes) using their API \citep{Swift0}.

A final important piece of infrastructure are data archives and repositories, especially those that directly store time-domain data.  These services are vital for data sharing and for collecting samples for statistical analysis. Particularly important for nearby SNe is WISeREP \citep[Weizmann Interactive Supernova Data Repository;][]{Yaron12}, which serves as an archive of SN spectra and photometry, where both new and published data can be stored.  Other important data repositories for nearby supernovae include the Open Supernova Catalog \citep{OSNCat} and OTTER \citep[the Open mulTiwavelength Transient Event Repository;][]{otter}.  Finally, the Mikulski Archive for Space Telescopes (MAST)\footnote{\url{https://archive.stsci.edu/}} is critical for SN progenitor studies when space-based imaging is available prior to explosion.

The time-domain infrastructure landscape is quickly evolving, and new tools are being built to coordinate observations or provide additional information to observers. One area that may deserve more innovation is science-focused brokers, which aggregate contextual and archival data for a given science case, perhaps with a tailored or filtered stream from the brokers.  Such an approach has been taken by the Solar System community \citep[e.g.,][]{Trilling23}.  Along these lines, we discuss the concept and opportunities for a Nearby Galaxy Transient Broker in Section~\ref{sec:discussion}.

\section{Supernova Overview}\label{sec:overview}

The period immediately surrounding a SN explosion -- from precursor emission to the hours after first light -- offers several observational clues to the explosion physics and progenitor system.  We summarize these signatures into broad categories in Table~\ref{tab:SNsignatures}.  For each, we present a representative absolute-magnitude range (in the optical) and the duration that the signature is visible.  These are meant as guides to help with survey planning rather than robust ranges, as each signature depends on numerous factors.  We also list typical distances out to which these signatures can be observed for three fiducial transient survey depths at $\sim 20$, 22.5, and 24 mag. Respectively, these are meant to approximately represent current programs (ZTF, ATLAS,  DLT40); imminent, deep programs to ``shadow'' LSST \citep{Ransomeshadow}; and the LSST WFD survey itself.  

The listed observational signatures are faint and short-lived.  Some features, such as the initial light-curve peak seen in Type IIb SNe, can be as bright as roughly $-17$ mag. However, the majority of signatures are $\gtrsim -14$ mag, a luminosity at which SNe are only rarely discovered. Fainter signatures are likely, but current surveys are not sensitive to such transients out to a large volume and/or do not have sufficiently high cadence to detect these fleeting features. 

We discuss these observational signatures in the following sections for core-collapse (Section~\ref{sec:ccsn}) and thermonuclear (Section~\ref{sec:thermo}) SNe.
As we discuss each signature, we also point out any infrastructure needs, required survey depths, and cadence requirements.  In order to estimate the number of early SNe that can be studied in each category, we make use of transient statistics from the TNS over the 5\,yr interval 2020--2024 inclusive, with a cut of declination $<$20 deg to mimic the WFD area, along with the distance and luminosity limits presented in Table~\ref{tab:SNsignatures}.  For nearby SNe ($\lesssim$200 Mpc), TNS statistics should  translate to approximate SN  numbers in the LSST era, as nearly all are followed up with at least one classification spectrum \citep[e.g.,][among others]{Fremling20,Perley20,Tucker22}.  The same may not be true for more-distant SNe, so there may be even more examples of bright early-time SN signatures than our estimates presented below.

%Provide overview plot, like the DLT40 plot.

%Paper goals: show science opportunities.  Point out what infrastructure/surveys will be vital, what is missing, etc.

\begin{figure*}
\centering
\includegraphics[width=0.49\textwidth]{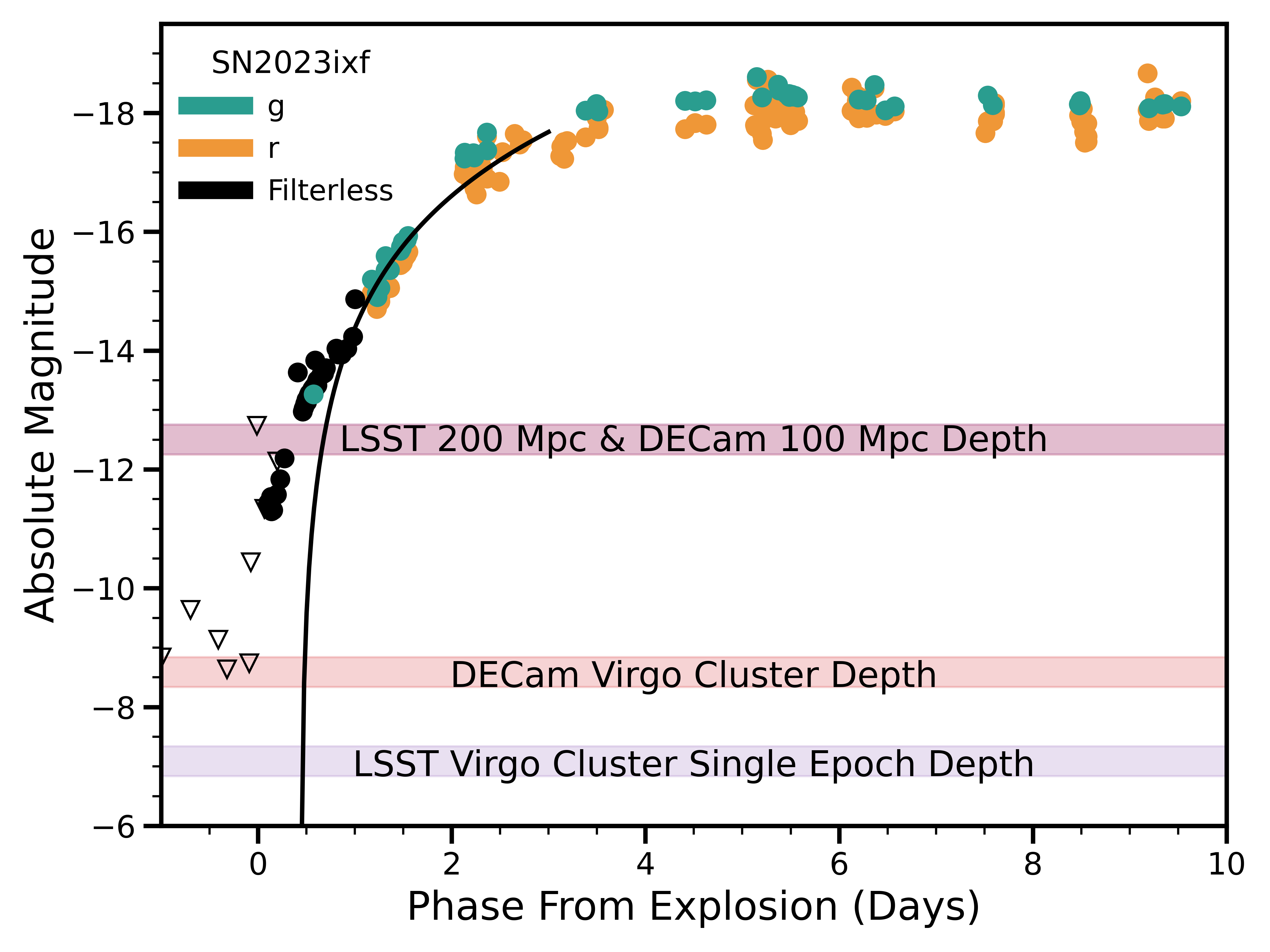}%\vspace{-20pt}
\includegraphics[width=0.49\textwidth]{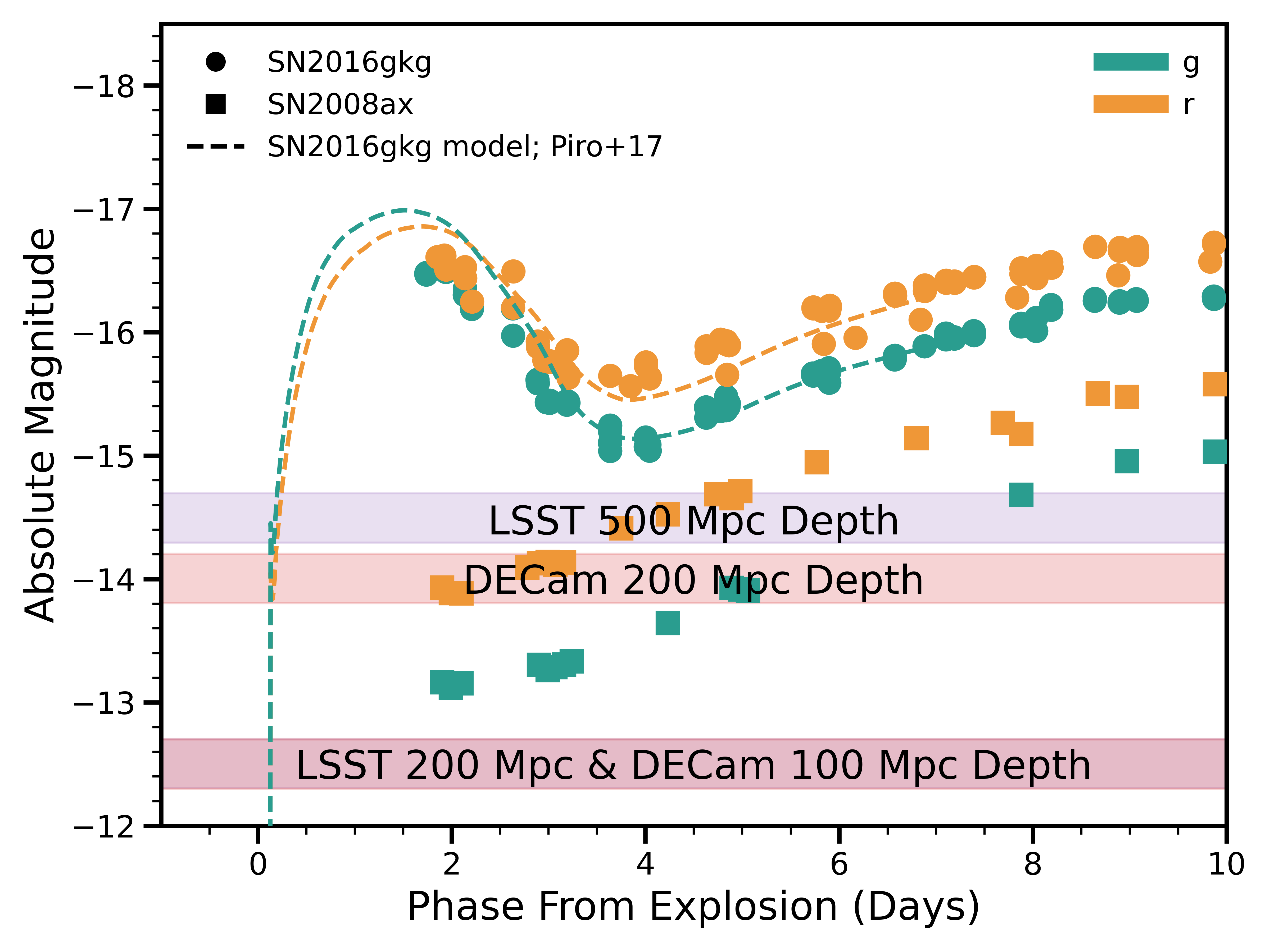}
\caption{Light curves of nearby core-collapse SNe which display early excesses indicative of shock-breakout cooling and/or early CSM interaction.  We also overplot the sensitivity of LSST and a nominal DECam shadowing program (Table~\ref{tab:surveys}). {\it Left:} Early light curve of SN\,2023ixf, a Type II SN in M101, including early amateur data (in black) and nondetections (upside-down triangles) \citep[from ][]{Hosseinzadeh23}.  An early excess was apparent after fitting the light curve 1--4 days after explosion and extrapolating backward in time (see solid-line quadratic fit from \citealt{Hosseinzadeh23}).  LSST will be able to detect such features out to $\sim$200 Mpc, but for sustained nightly cadence, a DECam shadowing program focused on $\lesssim$100 Mpc may be necessary. {\it Right:} Early-time light curves of two nearby SNe~ IIb: SN\,2016gkg \citep{Tartaglia17} and SN\,2008ax \citep{Pastorello08}.  We also include a numerical model of the light curve of SN\,2016gkg (dashed lines; \citealt{Piro17}), assuming an explosion within an extended envelope or CSM, as is expected for Type IIb SNe.  Early light-curve excesses like that seen in SN\,2016gkg will be discernible out to hundreds of Mpc with either LSST or a DECam shadowing program.  Weaker light-curve features, like that seen in SN\,2008ax, will be visible out to 200 Mpc (100 Mpc) with LSST (DECam shadowing).
\label{fig:CCSN_lightcurve}}
\end{figure*}

\begin{deluxetable*}{lcccccc}\label{tab:SNsignatures}
\tabletypesize{\footnotesize}
\tablecolumns{7}
\tablewidth{0pt}
\tablecaption{Early, Fast, or Faint Timescale Supernova Features}\label{tab:SNfeatures}
\tablehead{
\colhead{Supernova Feature} & \colhead{Timescale} & \colhead{Brightness} & \colhead{Probe of?} & \multicolumn{3}{c}{Distance (Mpc)}\\
\colhead{} & \colhead{} & \colhead{{(mag)}} & \colhead{} & \colhead{Shallow} & \colhead{Med/Shadow} & \colhead{LSST}\\
\colhead{} & \colhead{} & \colhead{{}} & \colhead{} & \colhead{$\approx$20 mag} & \colhead{$\approx$22.5 mag} & \colhead{$\approx$24 mag}} 
\startdata
\hline
Core-Collapse SNe \\
\hline
Shock-Breakout Cooling & $\lesssim$3 days & $-$17$<${}$M_{V}${}$<${}$-$11 & Progenitor Radius \& & $<$15-250 & $<$50-800 & $<$100-1600\\
&&& Energy/Ejecta Mass \\
&&& CSM Comp/Extent \\
Flash Spectroscopy & $\lesssim$3 days & $-$17$<${}$M_{V}${}$<${}$-$12 & CSM Comp/Extent & $<$25-250 & $<$80-800 & $<$160-1600\\
Spectropolarimetry & $\lesssim$2 days & $-$17$<${}$M_{V}${}$<${}$-$12 & Shock Breakout, CSM & \multicolumn{3}{c}{[instrumentation limited; $D\lesssim20$ Mpc]}\\
& & & \& Ejecta Geometry\\
Pre-Explosion Outburst & Days to years&  {}$M_{V}$$>$$-$15 & Massive Star Evol.  & $<$100 & $<$320 & $<$630\\
&&& Explosion Physics \\
&&& CSM Comp/Extent \\
~~~~~~~~~~Type IIn & & $-$15$<$$M_{V}$$<$$-$12 & & $<$25-100 & $<$80-320 & $<$160-630\\
~~~~~~~~~~Type Ibn & & $-$14$<$$M_{V}$$<$$-$11 & & $<$15-65 & $<$50-200 & $<$100-400\\
~~~~~~~~~~Type II & & $M_{V}$$>$$-$12 & & $<$25 & $<$80 & $<$160 \\
\hline
Thermonuclear SNe \\
\hline
Light-Curve Shocking & $<$5 days (RG) & $-$17$<$$M_{V}$$<$$-$16 &Nondegenerate  & $<$160-250 & $<$500-800 & $<$1000-1600\\
&$\lesssim$1 day (MS) &$-$16$<$$M_{V}$$<$$-$15& Companion \& CSM & $<$100-160 & $<$320-500 & $<$630-1000\\
Early Light-Curve Shape & $<$3 days & $M_{V}$$>$$-$14 & Nickel Dist. & $<$65 & $<$200 & $<$400\\
&&& Progenitor Radius \\
Carbon & Early as possible & NA & Expl Mechanism/Mixing \\
Very High Velocities & $\lesssim$3 day& $M_{V}$$>$$-$14 & Expl Mechanism/Mixing &$<$65 & $<$200 & $<$400 \\
($>$25,000 km s$^{-1}$)&&& Asymmetries/Shells? \\
\hline
\enddata
%\tablecomments{Yo}
%\vspace{-0.3in}
\end{deluxetable*}

\section{Core-Collapse Supernovae}\label{sec:ccsn}

Core collapse (CC) occurs in massive stars ($>$8 M$_{\odot}$) when fusion alone is unable to sustain the core against its own gravity.  CC~SNe are heterogeneous, comprising SNe IIP, IIL, IIb, IIn, and Ib/c \citep[e.g.][]{Filippenko97,Modjaz19}, with many hybrid variants --- all of which are the end states of massive stars with varying degrees of their envelope stripped at the time of explosion \citep[e.g.,][]{Heger03}. Open questions include the nature of the progenitor star system (e.g., over what
mass range do red supergiant stars explode as SNe II? what massive star or binary systems result in stripped-envelope SNe?) and the final years of massive-star evolution, as probed by the explosion interacting with circumstellar material (CSM). In the absence of rare pre-explosion space-based imaging with {\it HST} or {\it JWST}, the primary route to constrain the progenitor and its environment is through early-time observational signatures in the days following explosion --- or even the months to years preceding it --- as summarized in Table~\ref{tab:SNsignatures} and detailed below.

\subsection{Precursor Outbursts and Brightenings}\label{sec:precursor}

One probe of late-stage mass loss and instabilities in SN progenitors is the search for pre-explosion flaring or other precursor emission.  The final years of a massive star's life are still poorly understood.  There are several mechanisms that may lead to eruptive mass loss  and a concomitant electromagnetic brightening, including binary interaction \citep{Chevalier12}, turbulent convection in the core \citep{SmithArnett14}, and wave-driven transport \citep{Quataert12,Wu21}. %In some circumstances, a more exotic scenario may lead to `pre-cursor' emission, involving both a stellar merger and supernova that occur in the same system over the span of years \citep[e.g. SN2022mop;][]{Brennan25}.

Precursor light-curve brightening is seen in several subclasses of CC~SNe. We summarize the approximate brightness ranges for these events in Table~\ref{tab:SNfeatures} and Figure~\ref{fig:precursor}.  Precursor outbursts are especially common in SNe~IIn  \citep[e.g.,][]{Mauerhan13,Pastorello13,Ofek13,Ofek14,Tartaglia16,Strotjohann21,Hiramatsu24}, which are known to have extended, dense CSM \citep{Smith14}.  These outbursts have typical absolute magnitudes between about $-12$ and $-$15 mag, and may appear in $\sim$50\% of cases \citep[within 3 months of explosion;][]{Strotjohann21}, although the typical luminosity, color, evolution, and duration of such emission is still uncertain. Precursor emission is also seen in other CSM-interacting SNe, in particular the Type Ibn SN subclass \citep[e.g.,][]{Pastorello07,Strotjohann21,Dong24,Brennan24}, albeit at typically lower luminosities ($-11$ to $-$14 mag).  The precursor emission in the Type Ibn SN\,2023fyq was not ``eruptive,'' but instead displayed a steady brightening in the $\sim$150 days before the light-curve peak \citep{Dong24}. Finally, precursor emission has been seen in a single normal SN~IIP, SN\,2020tlf \citep{JG22,JG25}, starting $\sim$130 days prior to explosion, with an absolute magnitude of roughly $-12$ mag.  SN\,2020tlf displayed early flash features indicative of close, dense CSM which could be plausibly related to the precursor emission; however, other nearby SNe~II with flash features do not display any precursor emission to deeper limits (SN\,2023ixf, \citealt{Dong23,Ransome24,Panjkov24,Rest25}; SN2024ggi, \citealt{Shrestha24}; SN2024bch, \citealt{Andrews25}), suggesting that multiple physical mechanisms may be at play.

Given the nominal brightness of precursors, LSST's depth will revolutionize our view (Table~\ref{tab:SNsignatures},  Figure~\ref{fig:precursor}).  While previous surveys were sensitive to precursors within $\sim$10--100 Mpc, LSST will expand the accessible range out to $\sim$600 Mpc, depending on the precursor SN type and luminosity.  Precursors are extremely faint, although they tend to last for weeks or months \citep[but see][]{Davies22}, and thus the cadence of the LSST WFD survey is likely sufficient; greater depths and distances can be probed through data stacking over multiple epochs \citep[e.g.,][]{Ofek14}.

We can estimate how many opportunities LSST will provide for precursor measurements by taking transient statistics from the TNS, as described in Section~\ref{sec:overview}, combined with the distance sensitivity displayed in Table~\ref{tab:SNsignatures}.  There are $\sim$12--50 Type IIn SNe per year within $\sim$160--630 Mpc, suggesting that comparably sized samples will be available in the LSST era.  This is likely a lower limit, owing to the limitations of our TNS catalog (see Section~\ref{sec:overview}), and other work with more detailed modeling suggests that $\gtrsim$100 SN IIn precursors may be detectable per year \citep{Gagliano25}.  Similarly, we can expect $\gtrsim$5 Type Ibn SNe (at $\lesssim$400 Mpc) to be discovered every year.  Such numbers over the lifetime of LSST --- hundreds of SNe~IIn with potential precursor emission and dozens of SNe~Ibn --- will allow for precursor incidence rates to be calculated with high confidence, and the duration, color, evolution, and even spectral properties of such emission to be studied with statistical samples.

Equally interesting are the opportunities to detect precursor emission from normal Type II SNe, and extremely faint emission from all types.  If we use an absolute magnitude of $-$12 mag as a fiducial value for an SN~II outburst \citep[similar to that reported in SN\,2020tlf;][]{JG22,JG25}, then LSST will be sensitive to such emission out to at least $\sim$160 Mpc (and more if data are stacked, etc.), thereby sampling $\sim$150 normal SNe~II every year.  The fraction of normal SNe~II with SN\,2020tlf-like outburst emission will be known to $\lesssim$10\%  (based on counting statistics) within a few years of LSST operations.

Finally, the pre-explosion outburst limits on nearby SNe, like SN\,2023ixf, reached an absolute magnitude of $\sim -$9 with no detections (Figure~\ref{fig:precursor}, {\it right}; \citealt{Dong23}).  Outbursts at that luminosity and fainter are virtually unexplored territory owing to the volume and magnitude limits to which current surveys are sensitive.  LSST will vastly expand the volume of searches in this unexplored regime, as we illustrate in Figure~\ref{fig:precursor} ({\it right}).  Here we show that the LSST single-epoch depth at the distance of the Virgo Cluster ($\sim$18 Mpc) will reach absolute magnitudes of $\sim -7$.  Very nearby SNe will have unprecedented pre-explosion datasets from LSST.

There are several infrastructure tools and upgrades that will maximize the science that can be done with SN precursor emission in the LSST era.   One key goal will be to identify precursor emission as early as possible with transient broker filters associated with nearby galaxies so that they can be studied with multiwavelength facilities; even faint emission will be cause for follow-up observations, although the false-alarm rate will likely be high. Easy access to archival time-domain data from ongoing surveys (e.g., ATLAS, ZTF, LS4, BlackGEM) and earlier work (e.g., PTF/iPTF, the Catalina Real-Time Transient Survey) will be important to interlace with and extend any LSST pre-explosion detections.  Many of these surveys have longer temporal baselines than LSST and may provide decades-long coverage, a virtually unexplored regime in precursor work.  Along these same lines, access to astrometrically aligned high-resolution space-based imaging (e.g., HST, JWST) could identify a SN progenitor system prior to explosion, as the outburst is in progress.  %We also note that early in the survey transients may be detected by using archival DECam images as templates \citep[e.g. the SLIDE software package][]{Dong25}.

\subsection{Early Light Curves: Shock Breakout Cooling, CSM Interaction}\label{sec:ccsn_early_lc}

Several CC~SN subclasses can have initial light-curve peaks \citep[e.g.,][]{Tominaga05,Soderberg08,Mazzali08,Burke21} that may signal shock-breakout cooling or CSM interaction.  Here we first focus on the Type IIb subclass to illustrate the promise with LSST, as displayed in Figure~\ref{fig:CCSN_lightcurve} ({\it right}).  SNe~IIb are ``stripped-envelope SNe'' where part of the progenitor star's outer hydrogen and/or helium envelope has been shed.  In particular, SNe~IIb display weak hydrogen emission in their early-time spectra which give way to prominent helium lines after a few weeks \citep{Filippenko88,Filippenko97}, indicating a partially stripped envelope.

Many, but not all, SNe~IIb display a double-peaked light curve, with the early feature lasting up to $\sim$3--4 days after explosion, although some are more fleeting. 
This initial peak is likely due to shock-breakout cooling through the progenitor system's extended envelope or CSM.  If this feature is well-observed, models \citep[e.g.,][among others]{Sapir17,Piro21, Morag23} can yield strong constraints on the progenitor-star radius and envelope mass, which have been applied to several well-observed SNe~IIb with excellent data \citep[e.g.,][]{Arcavi_16gkg,Tartaglia17,Bersten18,Armstrong21,Pellegrino23,Das23,Farah25,Subrayan25,Franz25,Yete26}. By mapping out these quantities in the SN~IIb population, we can constrain the  envelope-stripping mechanism \citep[e.g., stellar winds, binary interactions, rotational stripping; see][for a review]{Smith14} and ultimate origins of these explosions.

To illustrate the capabilities of LSST and a DECam-like shadowing campaign in particular, we plot the early $g$- and $r$-band light curves of two Type IIb SNe (Figure~\ref{fig:CCSN_lightcurve}, {\it right}): SN\,2016gkg \citep{Tartaglia17} and SN\,2008ax \citep{Pastorello08}.  These two SNe have high-cadence early light curves, and show the range of signatures in this subclass --- on one end, SN\,2016gkg displays a prominent early-time excess that lasts for $\sim$3--4 days, while SN\,2008ax is overall fainter and shows a very mild excess (although it could have been more prominent $\lesssim$1 day after explosion).  The LSST WFD will have sensitivity to the early light-curve excess seen in SN\,2016gkg out to $\gtrsim$500 Mpc, while a DECam shadowing campaign ($\sim$22.5 mag depth) could do so beyond $\gtrsim$200 Mpc. These numbers are closer to $\gtrsim$200 Mpc ($\gtrsim$100 Mpc) for identifying any early light-curve features in SN\,2008ax-like SNe with LSST (DECam).  Based on our TNS statistical sample, this implies a possible set of $\gtrsim$5 and $\gtrsim$15 SNe per year for SN\,2008ax-like and SN\,2016gkg-like cases, respectively, for the conservative DECam shadowing case.

The opportunity to observe early, subtle features seen only once or twice before will also increase. The initial detections of the nearby Type II SN\,2023ixf served as a preview of the LSST era, as we illustrate in the left panel of Figure~\ref{fig:CCSN_lightcurve}.  This SN exploded in the nearby galaxy M101 ($D \approx 6.7$ Mpc), discovered by an amateur astronomer \citep{SN23ixf_discover}.  The community response was immediate, and since M101 is a prominent amateur target, pre-discovery imaging by that community played an important role in tracing the light curve in the hours after explosion \citep{Sgro23}.  With this extremely early amateur data, \citet{Hosseinzadeh23} identified an early excess in comparison to the later rise of the light curve after $\sim$0.5d; further multiband pre-discovery data were collected by \citet{Li24}.  While the origin of this excess is unclear, initial interpretations have pointed to CSM interaction and/or shock-breakout cooling from a dusty red supergiant (RSG) as plausible culprits.  The LSST WFD on its own could detect early light-curve excesses similar to that seen in SN\,2023ixf out to  $\sim$200 Mpc.  However, it is likely that a shadowing campaign (coupled with intensive follow-up observations) is necessary to detect such fleeting signals (which last $\lesssim$1 day), where a DECam program could see such features out to $\sim$100 Mpc.  Within this accessible volume and the LSST WFD footprint there are $\sim$75 normal SNe~II per year, suggesting that potentially hundreds of SN\,2023ixf-like excesses could be searched for and studied with semi-dedicated shadowing efforts over the course of LSST's lifetime.  Clearly, with only one SN\,2023ixf-like example, %basic questions about their physical origin remain unanswered, and 
more data is needed to constrain the incidence, color, and morphology of excesses at the earliest times.  
%A sample of similar early light curve data is needed to constrain the incidence, color, and morphology at the earliest times, and begin to address basic questions about their physical origin.  %Further examples are necessar will begin to shed light on their physical origins.

Importantly, while the sensitivity of the WFD survey will open up a much larger volume for identifying early light-curve excesses, its cadence may make identifying such features very difficult without shadowing efforts and/or complementary surveys to identify young objects.  Sustained, high-cadence follow-up observations with multiband facilities is also vital.  
The needs for observing early light-curve signatures are demanding, as these features last only hours to days, and they are faint, $\gtrsim$$-$15 mag.  Early identification of the SN candidate is vital so that multiband monitoring can begin within hours and maintained over several days while features develop and the spectral energy distribution of the SN quickly evolves.  Few facilities or SN collaborations have access to the instrumentation necessary for such a campaign.  If the SN is nearby, Las Cumbres Observatory's network of 0.4\,m to 2\,m  telescopes around the globe is ideally suited for the task \citep{Brown13}.  Otherwise, SN teams (and amateurs) with access to telescope facilities (and associated fast reaction infrastructure, such as TOMs with automated filters and telescope API access) across different longitudes should band together to produce continuous high-cadence light curves.

%Need paragraph on order of magnitude expected.

\subsection{Early Spectroscopy: Flash Features and their Incidence}
%\subsection{Early Light Curves: CSM Interaction}
A significant fraction of CC~SNe observed within a few days of explosion exhibit narrow emission lines corresponding to highly ionized species \citep[e.g.,][]{khazov16,bruch21,bruch23}.  This emission is likely a signpost of dense, confined CSM around the progenitor and is caused by either shock-breakout ionization \citep[e.g.,][]{Yaron17} or ejecta interaction  \citep[e.g.,][]{Terreran22}, and it lasts only a few days before standard SN~II features take over.  Detailed observations and modeling of these ``flash'' features can constrain the mass-loss rate of the progenitor just prior to explosion, the surface chemical composition, and the extent of this confined, dense CSM \citep{Dessart17,Boian19,Boian20}. The origin of this material is unclear (as we discussed in Section~\ref{sec:precursor}), as it far exceeds that expected from standard RSG winds \citep[see further discussions by][]{Kochanek19,Davies22}.  

%Perhaps the best way to understand the mass loss history of supernova progenitors in the years leading up to explosionEven closer to explosion is necessary.  Do all CC SNe show flash features?  Do those with the early ledge show flash features if observed early enough?

Although dozens of examples of the flash phenomena have been published \citep[e.g.,][]{Quimby07,Galyam14,Smith15,Yaron17,khazov16,bruch21,bruch23,JG22,JG23,JG24,JG24_SN24ggi,Tartaglia21,Terreran22,Bostroem23,Shrestha24,Andrews24,Andrews25,Ransome26}, the time evolution and duration of the emission lines is still not well-constrained owing to the ephemeral nature of the emission \citep[although see][]{JG24}.  In the best-observed systems, clear evolution proceeds over timescales of hours \citep[e.g.,][]{Bostroem23,Shrestha24,Pessi24}, suggesting that both fast response and sustained monitoring is necessary to fully characterize this phase.

A subset of early CC~SNe do not exhibit clear narrow emission lines, but instead display broadened, blueshifted emission around $\sim$4600\,\AA, sometimes referred to as a ``ledge'' feature \citep[e.g.,][]{Bullivant18,Andrews19,Soumagnac20,Hosseinzadeh_21yja,Pearson23,Nico24,Shrestha_MMT,Lin25,Dubey26,Gerard26,Ailawadhi26}.  This feature may be associated with low-density CSM, such that the narrow emission lines do not develop, or are only visible for a very short period of time prior to the broadened emission \citep[e.g.,][]{Dessart17}.  This broadened feature is also apparent as narrow emission lines fade in other SNe.  It is possible that all ledge SNe exhibit narrow emission lines if spectra are obtained sufficiently early, although this is an open question.

Another frontier area is high-resolution spectroscopy during the flash phase, where the details of the emission lines can be studied, possibly leading to clearer signatures of the progenitor star.  Thus far, such early-time flash spectra have  been obtained only for SN\,2023ixf \citep{Smith23_highres,Dickinson25}, SN\,2024ggi \citep{Pessi24,Shrestha24},
 SN\,2024bch \citep{Andrews25}, and SN\,2025ngs \citep{Ransome26} (as well as the Type IIn SN\,1998S, which had longer-lasting emission lines than the objects we discuss here; \citealt{Shivvers15}), but features in these spectra point to kinematic signatures that are $\sim$10--100 km s$^{-1}$, suggestive of RSG winds.  There is also evidence for acceleration of the dense CSM material \citep{Dickinson25} and rapid evolution of the \ion{C}{4} line profile (which goes from double-peaked to single-peaked in $\sim$7 hr; \citealt{Shrestha24,Pessi24}).  In SN\,2025ngs, the appearance of a double-horned H$\alpha$ profile with width $\sim$40 km s$^{-1}$ suggested a disk-like CSM morphology, which only lasted for one day \citep{Ransome26}. Further data and modeling is necessary to understand the physical implications of these observations.

Flash-spectroscopy features have been seen in CC~SNe at a range of luminosities, $M_V > -17$ mag and fainter.  At this luminosity, LSST will be sensitive to such SNe well beyond 1\,Gpc (and thousands of SNe per year), although this is beside the point, as the most distant examples will not be amenable to follow-up spectroscopy.  If instead we consider the more practical limit of a shadowing survey ($\sim$22.5 mag) that focuses on nearby galaxies within $\lesssim$160 Mpc (corresponding to the distance at which LSST will be sensitive to SN\,2020tlf-like pre-explosion outbursts), there will be $\sim$150 normal Type II SNe every year.  A dedicated program, even without 100\% coverage but with sufficient statistics, would be able to constrain the fraction of SNe~II with flash  features to $\lesssim$10\% within a couple of years of LSST operations.  The collection of further high-resolution spectra of nearby, young CC~SNe will largely be limited by current telescopes and instrumentation, where spectra having high signal-to-noise ratio (S/N) require $V \lesssim 17$ mag, limiting targets to a handful per year where they can be studied from the earliest moments.

The infrastructure needs for early spectroscopy of normal CC~SNe require fast reaction and sustained follow-up observations. Similar to the light-curve science case, extremely early identification is necessary to get definitive statistics on the fraction of CC~SNe that display early flash features, as improvement will come by shrinking the time between explosion and the first science-grade spectrum.  From there, the evolution of flash features (on timescales of hours) will provide details on the density profile and extent of the confined material.   This again requires worldwide observations and cooperation in obtaining spectroscopic sequences --- something that large transient collaborations with strong infrastructure (including TOMs with telescope API hooks to trigger across facilities) can accomplish.

%Related: ledge feature.  if you looked early enough, would there be flash features?  if you looks early enough, would they all have flash featurs?

%Multi-epoch sequences.

%Frontier is high resolution.

\subsection{Early Polarimetric Observations} 
Early light curves and spectral time series can teach us about shock breakout and early CSM interaction, but spectropolarimetry can play a special complementary role in probing the {\it three-dimensional geometry} of the ejecta and CSM.  Continuum polarization measures global departures of the photosphere from spherical symmetry, while line polarization traces the distribution of specific chemical species in the ejecta. Knowledge about the symmetry, or lack thereof, of an SN explosion and the material it interacts with provides fundamental information \citep[for a review, see][]{Wang08}.

The power of early-time spectropolarimetry has been vividly demonstrated by two recent, nearby CC~SNe, SN\,2023ixf and SN\,2024ggi.  Both were discovered early, and follow-up observations were obtained within days of explosion, providing glimpses of the future LSST era. In the case of SN\,2023ixf, three independent spectropolarimetric datasets spanning +1 to +120 days after explosion revealed a rich, evolving geometric picture \citep{Vasylyev23,Vasylyev25,Singh24,Shrestha25}. During the first few days, when flash-ionization features were prominent and the light curve was rising, the intrinsic continuum polarization reached $\sim 1\%$ --- the highest level detected at such an early time for any normal SN~II --- indicating that the surrounding CSM was aspherical. Radiative-transfer modeling of this early polarization requires CSM with a pole-to-equator density contrast of $\gtrsim 3$ \citep{Vasylyev25}. The polarization then declined steeply through the photospheric plateau phase as the expanding photosphere swept up the optically thick CSM and approached spherical symmetry, only to rise again to $\sim 0.5$\% near the fall from plateau, attributed to either an aspherical distribution of 
$^{56}$Ni deep in the ejecta \citep{Vasylyev25} or an aspherical helium core \citep{Shrestha25}. Inverse P~Cygni polarization profiles observed in H and He lines during the plateau phase further indicate asymmetrically distributed intervening CSM overlying the photosphere. Together, this temporal sequence reveals distinct sources of geometric asymmetry at early and late times: an aspherical, confined CSM close to the progenitor, and an asymmetric inner ejecta structure accessible only after the photosphere recedes.

SN\,2024ggi provided an even earlier window into the explosion geometry \citep{Yang25}.  VLT/FORS2 spectropolarimetry beginning just $\sim$1.1 days after explosion  revealed a well-defined dominant axis in the Stokes $Q-U$ plane --- a signature of an axially symmetric shock-breakout geometry.  In the days that followed, the continuum polarization changed dramatically, suggesting emission that cannot arise from a single axisymmetric structure. \citet{Yang25} conclude that the shock breakout and the surrounding CSM have distinct, misaligned symmetry axes, requiring aspherical ejecta breaking out into disk-concentrated CSM. Such recovery of the geometry of the explosion and the surrounding CSM (a remnant of the pre-explosion star) can only be gathered by very early, persistent polarimetry.

These two events illustrate the richness of early-time polarimetric signatures in CC~SNe: close-in CSM orientation, shock-breakout geometry, inner-ejecta asymmetries, and the connection to the explosion mechanism itself are all potentially encoded.  The key observational requirement is obtaining spectropolarimetry within $\lesssim$2 days of explosion, before any flash features fade and before the photosphere circularizes, which demands both rapid detection and fast follow-up response. This is extremely demanding in practice: high-S/N  spectropolarimetry is currently feasible only for targets reaching $V\lesssim17$ mag, limiting this technique to only the brightest SNe in the sky caught at such early phases.

\section{Thermonuclear Supernovae}\label{sec:thermo}

Type Ia SNe play a starring role in cosmology, from the discovery of the accelerating expansion of the Universe driven by dark energy \citep[][]{Riess98,Perlmutter99} to the recent ``crisis'' in the measurement of the Hubble constant derived locally \citep[e.g.,][]{Riess22} and that inferred from the cosmic microwave background radiation \citep{Planck18}.  Despite this importance, a deep understanding of SN Ia progenitors and explosion physics is still lacking \citep[see][for a review]{Jha19}.  While it is accepted that SNe Ia are the thermonuclear explosions of carbon-oxygen white dwarfs in binary systems, fundamental questions remain as to the nature of the companion (e.g., single or double degenerate), the mass of the primary white dwarf (Chandrasekhar mass $M_{\rm Ch}$, or not), and the explosion mechanism (deflagration, violent merger, delayed detonation, double detonation, etc.).  There are also other types of peculiar white-dwarf explosions (Type Iax, 02es-like, ``Ca-rich'' transients, ``super-Chandrasekhar'' or SN\,2003fg-like, SN Ia-CSM, etc.), which shed light on plausible white-dwarf progenitor configurations and how they might explode \citep[see ][for a review]{Taubenberger17}.
Several early-time observational signatures can shed light on the physical picture of thermonuclear SNe, which we discuss next, along with the opportunities in the LSST era.

\subsection{Early Light-curve Excesses}

One of the most promising avenues to understand thermonuclear explosions are through their early light curves ($\lesssim$1--3 days from first light), which can constrain the radius of the progenitor, the distribution of $^{56}$Ni, and the presence of a normal (nondegenerate) companion star or CSM.  We list the range of luminosities and timescales associated with these mechanisms in Table~\ref{tab:SNsignatures}, and discuss them further below. Several normal SNe~Ia  have had  excesses or bumps in their very early light curves (within the first few days) which have been both blue and red in color \citep{Marion16,Hosseinzadeh17,Hosseinzadeh22,Hosseinzadeh_23bee,Dimitriadis19_18oh,Stritzinger18,Shappee19,Ni22,Wang24,Hoogendam24,Pearson24,Burke25}.  In addition to normal SNe Ia, two thermonuclear classes appear to commonly have early light-curve excesses or nonmonotonic rises: the subluminous, SN\,2002es-like subtype \citep{Cao15,Miller20,Burke21,Srivastav23} and the carbon and oxygen-strong SN\,2003fg-like objects \citep{Jiang21,Dimitriadis23,Srivastav22ilv}.    

First, the radius of the progenitor can be constrained by a well-sampled early light curve and the limits on any shock-breakout emission.  Such work has yielded a radius constraint of $R \lesssim 0.02$ $R_{\odot}$ for SN\,2011fe \citep{Bloom12}, and less stringent constraints in other systems \citep[SN\,2013dy; ][]{Zheng13}.  The SN\,2011fe observations in particular demand that a degenerate object is the progenitor of this SN~Ia, empirical evidence for the widely accepted view for normal SNe~Ia. To make further, nearly direct constraints on shock-breakout emission in SNe~Ia requires nearly immediate discovery and monitoring, which will remain rare in the LSST era, although modern time-domain infrastructure (e.g., automated follow-up observations of nearby galaxy transients from filtered broker streams) may enable this.

The single-degenerate scenario predicts that the SN~Ia ejecta collision with the secondary, normal star will produce a shock visible at high energies all the way into the optical, depending on the viewing angle \citep{Kasen10}.  The optical signal is only visible for a day or two for a main-sequence companion, and only $\sim$10\% of the time is an observer at the right viewing angle. The shock luminosity is $-17<M_V<-15$ mag depending on the companion type \citep{Kasen10}, requiring both deep and high-cadence observations to ensure early capture of the signal.   Similarly, close-in CSM (not necessarily from a nondegenerate companion) can also have effects on the very early SN~Ia light curve \citep{Piro16}, but visible only on $\lesssim$2 day timescales.

Aside from shock breakout or interaction with a companion star (or CSM), the first photons that leave the SN ejecta are from energy deposition from the outermost layers of $^{56}$Ni, or other radioactive isotopes depending on the situation (see below).  As the ejecta expand, photons generated by energy deposited by deeper $^{56}$Ni escapes.  Thus, the early light curve could have a different shape depending on the radioactive isotope distribution \citep[e.g.,][]{Piro13}.  This distribution is expected to vary between explosion models and white-dwarf progenitors, making early light curves an important probe.  For example, an unusual distribution of $^{56}$Ni, coming from shells or clumps near the surface of the ejecta, can cause a blue early excess \citep{Magee20,Magee20b}.
Double detonations that start with surface helium fusion may have an early light-curve excess or bump that may be reddened owing to the production of Ti and Ca \citep[e.g.,][]{Jiang17,Ni23}.

\begin{figure}
    \centering
    \includegraphics[width=0.47\textwidth]{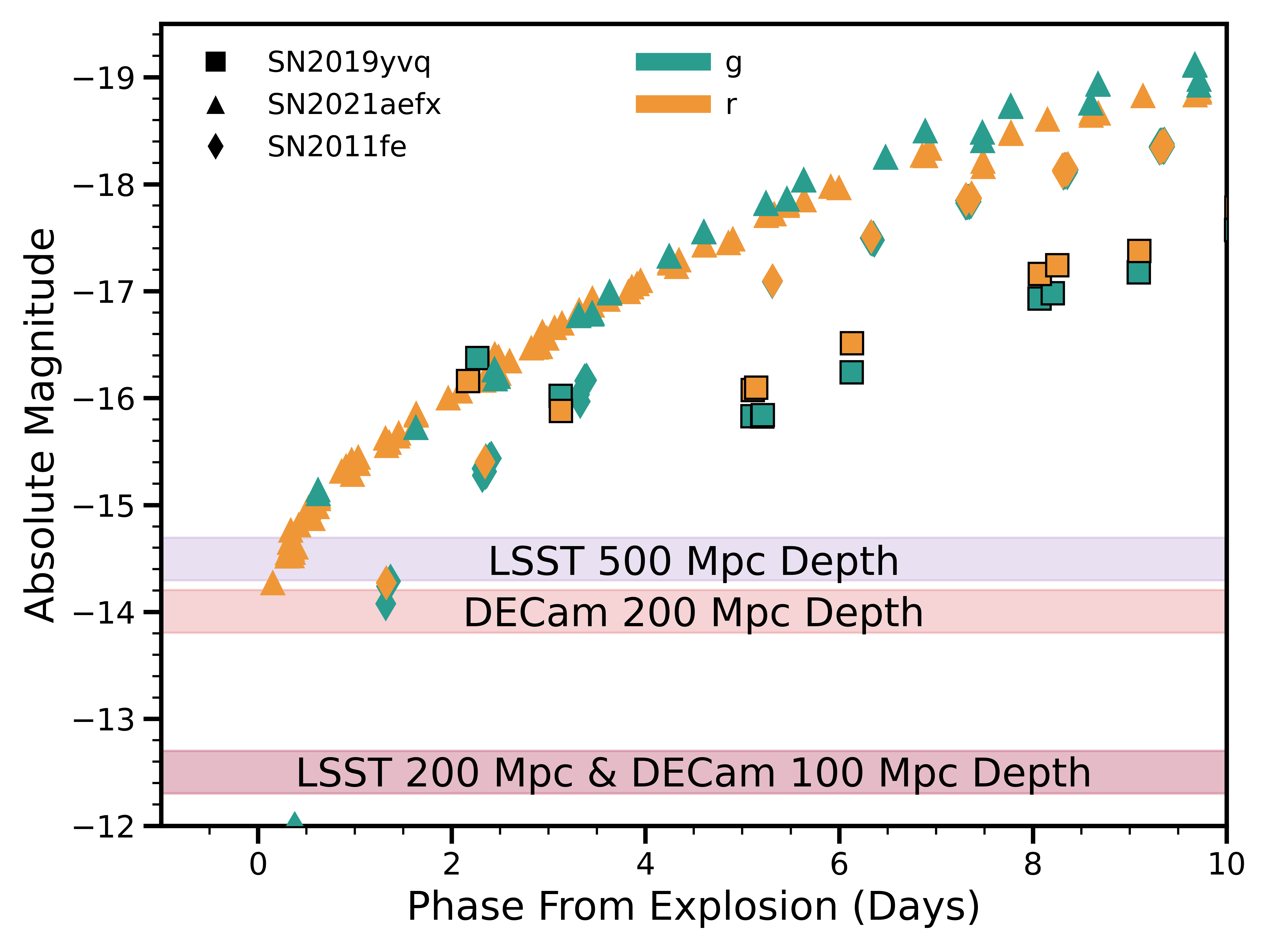}
    \caption{Light curves of two nearby thermonuclear SNe with early light-curve excesses: SN\,2019yvq, an underluminous, 02es-like SN Ia \citep{Burke21}, and the normal SN Ia 2021aefx, which also showed extreme high-velocity features \citep{Hosseinzadeh22}.  We also overplot the early light curve of SN\,2011fe \citep{Nugent11,Firth15,Graham17}, which did not have an early light curve excess.  Alongside these data, we overplot the sensitivity of LSST and a DECam shadowing program at various distances.  A joint LSST+DECam campaign can capture these and fainter light-curve features out to $\sim$200 Mpc provided high-cadence photometric monitoring is conducted.}
    \label{fig:thermo}
\end{figure}

We illustrate the opportunity for observing early thermonuclear light-curve features with LSST and shadowing programs in Figure~\ref{fig:thermo}.  Here we have plotted the early $g$- and $r$-band light curves of the normal Type Ia SN\,2021aefx \citep{Hosseinzadeh22} and the underluminous 02es-like SN\,2019yvq \citep{Burke21}, both of which showed excess emission in their early data (for SN\,2019yvq, this light-curve excess was even more evident in the UV).  Given the luminosity range of such features ($M_V \gtrsim -14$ mag; see Table~\ref{tab:SNsignatures}), these should be visible out to $\sim$400--500 Mpc for LSST, or $\sim$200 Mpc for a carefully planned shadowing survey.  Within that $\lesssim$200 Mpc, there are $\sim$250 SNe~Ia classified per year, without accounting for the numerous thermonuclear cousins that have now been uncovered (such as SN\,2019yvq in Figure~\ref{fig:thermo}).  The fraction of SNe~Ia with early light-curve excesses is still poorly constrained (although it could be $\gtrsim$10\%; \citealt{Burke25}), but during the lifetime of LSST it is possible that hundreds of thermonuclear light-curve excesses will be identified and the fraction of SNe~Ia displaying such features will become known to a few percent (along with their color, duration, morphology, etc.), strongly constraining shock interaction and other models for their origins.

Despite the opportunity of finding hundreds of thermonuclear SNe with early light-curve excesses, gathering such data will be extremely challenging.  The circumstance is similar to the early CC~SN light-curve case discussed in Section~\ref{sec:ccsn_early_lc}, where world-wide collaborations that share telescope resources are necessary, alongside state-of-the-art time-domain infrastructure.

\subsection{Early Spectroscopic Features}

A handful of SNe Ia have now been discovered so young that their earliest spectra reveal unprecedentedly high velocities, $\gtrsim$ 25{,}000--30{,}000 km~s$^{-1}$, evidence that the photosphere at this phase probes the outermost layers of the ejecta.  As the most extreme example, SN\,2021aefx showed high-velocity features within the first two days when the Si\,{\sc ii} $\lambda$6355
velocity exceeded $\sim$ 30{,}000
km s$^{-1}$ \citep{Hosseinzadeh22}. Similarly extreme silicon velocities were seen in SN\,2017cbv and SN\,2023bee at early phases \citep{Hosseinzadeh17,Hosseinzadeh_23bee}; we show the early spectrum of SN\,2023bee in Figure~\ref{fig:templatemismatch} and discuss related spectral classification issues with these early spectra in Section~\ref{sec:classify}.  It is worth noting that all three of these SNe also had early light-curve excesses, although the sample size is too small to suggest any definitive correlations.  Other SNe Ia with very early spectroscopy, like SN\,2011fe, did not exhibit such high silicon velocity features \citep{Parrent12}, suggesting that these signatures are not ubiquitous.  Currently, only a handful of such events exist, drawn from a narrow corner of SN Ia parameter space, and it is impossible to assess whether these high velocities are common across the full range of SN Ia decline rates and subtypes. Systematic early spectroscopic coverage of a large SN Ia sample, as enabled by shadowing the LSST WFD survey, is essential to make progress.

The physical origin of such high-velocity material remains unclear; proposed explanations include an enhanced density in the outermost ejecta from the explosion itself from blobs or fingers of clumpy material \citep{Mazzali05,Tanaka06}, interaction with a compact circumstellar shell \citep{Mulligan17,Mulligan19}, and viewing-angle-dependent signatures of double-detonation models \citep[e.g.,][]{Boos24}. 
No matter the mechanism, these early high-velocity features carry diagnostic power for the explosion physics and/or outermost ejecta structure that is lost once the photosphere recedes inward.

Other very early spectroscopic signals encode information about thermonuclear explosions. Carbon absorption features detected in early SN Ia spectra provide a direct probe of unburned material and thus a powerful discriminant between explosion models. Carbon is expected to be present in the progenitor C/O white dwarf but should be consumed in the explosion; its survival in the outer ejecta can indicate incomplete burning, off-center or asymmetric detonation, or specific ignition conditions tied to the explosion mechanism  \citep[e.g.,][]{Hoeflich95,Hoeflich96,Hoeflich02,Gamezo03,Gamezo04,Ropke07,Pakmor12,Shen14,Dessart14}.

Testing for carbon requires spectra quickly after explosion, and the earlier data are taken, the more likely carbon is to be found.  In the optical, $\sim$20-40\% of early-time spectra show C\,{\sc ii} $\lambda$6580 features \citep[e.g.,][]{Parrent11,Thomas11,Blondin12,Folatelli12,Silverman12,Maguire14,Wyatt21}, which appears on the red
shoulder of the Si\,{\sc ii} $\lambda$6355 absorption line.  While these features are typically weak and fleeting in normal SNe~Ia, they are a ubiquitous defining characteristic of super-Chandrasekhar (03fg-like) SNe~Ia, where strong C\,{\sc ii} absorption often persists past peak brightness \citep[e.g.,][]{Howell06, Ashall21, Dimitriadis22}, again emphasizing that the presence of carbon likely signals something fundamental about the explosion or progenitor system.

Moving beyond optical spectra is also necessary for a full accounting of carbon in SNe Ia, as the near-infrared (NIR) C\,{\sc i} $\lambda$10693 line is often detectable.  For normal SNe Ia, there have been several claims of a C\,{\sc i} $\lambda$10693 detection, manifesting as a ``shoulder'' in the more prominent Mg\,{\sc ii} $\lambda$10927 line that gets stronger toward maximum light \citep{Hsiao13,Hsiao19,Marion15}.  Even more intriguing, subluminous SNe Ia seem to display a different C\,{\sc i} $\lambda$10693 morphology, with strong  and distinct absorption clearly visible at early times, followed by a decline of the feature toward maximum light \citep{Hsiao13,Wyatt21,Pearson24}, although this absorption was seen to persist even longer in SN\,1999by \citep{Hoeflich02}.  Some studies have claimed this feature is actually He\,{\sc i} $\lambda$10830, rather than C\,{\sc i}, and that it is a signature of the double-detonation scenario in these subluminous SNe Ia \citep{Boyle17,Collins23}.  Either way, early-time NIR spectroscopy of SNe Ia has the potential to shed light on the explosion mechanism, and should be an integral part of any nearby SN campaign.

%A couple of recent type Ia supernovae have been discovered so young that their initial spectra show unprecedentedly high velocities ($\sim$30,000 km s$^{-1}$, or $\sim$0.1$c$), direct evidence that the photosphere at this phase is probing the outermost layers of the ejecta.

\section{Key Challenges} \label{sec:discussion}

In this work we have highlighted scientific opportunities in the LSST era for obtaining exemplary datasets on the nearest and brightest SNe, which can shed light on their progenitors and explosions.  These opportunities will be enabled by the numerous time-domain surveys which are shadowing the WFD survey, producing an effective cadence much less than the nominal $\sim$3 day cadence of LSST.  Here we discuss challenges for shadowing projects focused on nearby SNe, some of which are astrophysical in nature and others of which are technical. %to this idea, some of which are inherent to the nature of young supernovae, and some of which are technical, but can be overcome.

\subsection{Early Spectral Classification Ambiguity}\label{sec:classify}

In the hours after explosion, SN signatures can be ambiguous, and young transients are often misclassified or inconclusive.  This can happen because SNe at these phases display their highest velocities or are dominated by hot blackbody-like emission, making their classification ambiguous until lines develop.

\begin{figure}
    \centering
    \includegraphics[width=0.47\textwidth]{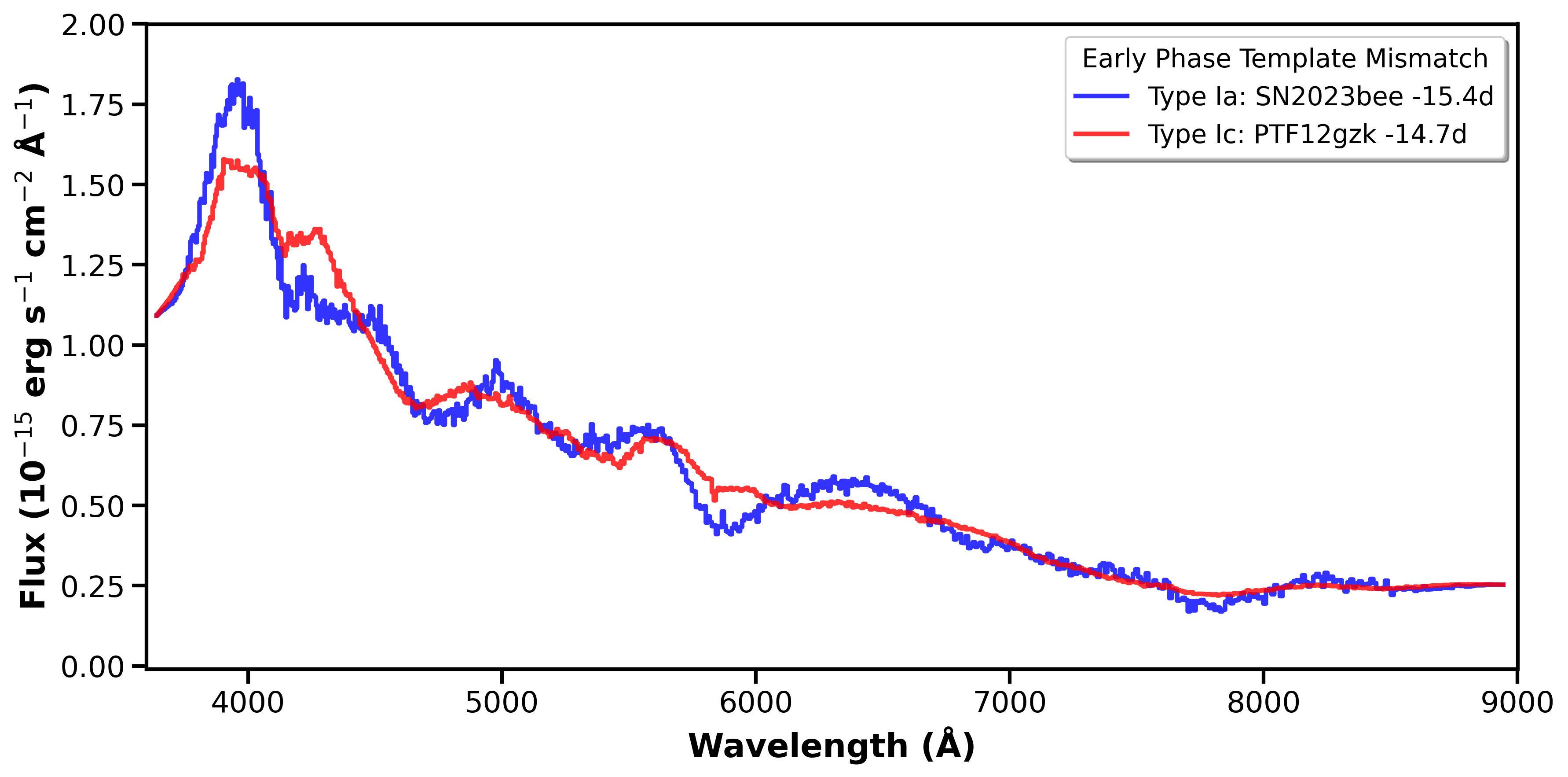}
    \caption{The very early spectrum of SN\,2023bee \citep{SN23bee_class_Ic} displayed an extremely high $v_{\rm Si\,{\sc ii}}=25,220$ km s$^{-1}$, leading to an early misclassification as an SN Ic, similar to the match to the Type Ic PTF12gzk \citep{benami12} shown here.  A spectrum taken the following day established SN\,2023bee as a Type Ia SN \citep{sn23bee_classify}.}
    \label{fig:templatemismatch}
\end{figure}

Early spectral classification uncertainty is possible for even the most common types of SNe.  For instance, the very earliest spectrum of the Type Ia SN\,2023bee resulted in an initial classification as a Type Ic SN \citep[see Figure~\ref{fig:templatemismatch};][]{SN23bee_class_Ic}, largely because of the extreme velocity at this epoch \citep[$>$25,000 km s$^{-1}$;][]{Hosseinzadeh_23bee}.  Similarly, the early spectrum of SN\,2021aefx was unclear, so it was initially classified as simply a Type I SN \citep{SN21aefx_initialclassify} before a spectrum taken a day later provided a definitive SN Ia classification \citep{SN21aefx_classify}.

Some SN types are not distinguishable until days or weeks after explosion.  For instance, stripped-envelope SNe are most distinguishable  $\sim$10--15 days after explosion \citep[e.g.,][but see \citealt{Williamson23,Yesmin25}]{Williamson19}, while others are truly ``transitional'' objects between subclasses that shed light on SN origins \citep[e.g.,][]{Dong24_22crv}.  The class of ``calcium-strong transients,'' which may be stripped-envelope SNe or white-dwarf explosions, first present hot blackbody spectra and then undergo a rapid transition to the nebular phase with strong calcium that finally distinguishes them; their definitive classification is often delayed because of this \citep[e.g.,][]{JG22,Chen26,Ravi26}.

To mitigate early ambiguity, new template spectra should be generated from the very youngest SNe and added to existing classification software \citep[e.g.,][]{superfit,snid,ngsf,snid_sage_2025}; such efforts are already underway \cite[e.g.,][]{Yesmin25}. %These should minimize confusion although not eliminate it, as some SN types are not distinguishable until days or weeks after explosion \citep[e.g.][]{Williamson19}, while others are truly `transitional' objects between subclasses that shed light on SN origins \citep[e.g.][]{Dong24_22crv}.  
In the end, intensive datasets of nearby, young transients should be collected by the large SN collaborations (and be made public; e.g., \citealt{passta}) regardless of type so that no matter what the ultimate classification, data are available. 

\subsection{Fast-response Space-based Observations}

In this paper, we have primarily focused on the ground-based ecosystem for the discovery and follow-up observations of nearby SNe.  However, space-based facilities have a huge role to play. For instance, X-ray and UV observations are especially sensitive to shock breakout and CSM interaction, while the early IR properties of most transients are still unexplored.  Space-based imaging prior to explosion can sometimes be used to identify the progenitor star and its environment. 

First, space {\it survey} facilities will have a large impact and will require prompt follow-up observations akin to those discussed in Sections~\ref{sec:ccsn} and \ref{sec:thermo}.  The wide field (3600 deg$^2$) and soft X-ray sensitivity of the Einstein Probe \citep{Yuan22,Yuan25} can detect nearby shock breakouts \citep{oconnor26,suzuki26,MC26,Wen26,Rastinejad26} and is an excellent complement to UV and ground-based searches for young SNe.  Similarly, ULTRASAT \citep{ULTRASAT} will identify shock breakouts of nearby CC~SNe via its high-cadence UV observations of $\sim$200 deg$^2$, potentially tens per year.  Even with this strong discovery capability, ULTRASAT has only a single UV filter and will need fast follow-up observations across the electromagnetic spectrum to characterize its discoveries.  The Roman Space Telescope \citep{roman} will be a premier space time-domain survey facility, but will largely be focused on the high-redshift universe unless a nearby-transient program is accepted as part of its General Astrophysics Surveys program; Roman would be ideal for identifying precursor activity, obscured SNe, and a zoo of dusty and interacting transients. %future General Astrophysics Surveys with many high priority %to provide(and many other UV transientsCurrent JWST and Roman transient surveys are largely focused on the high redshift universe, but searches for nearby transients with either facility would be quite powerful.

%Here we focus on early time space-based observations, within a few days of explosion.  
Opportunities for space-based {\it follow-up} observations within a few days of explosion are currently limited, but there are promising opportunities on the horizon. The Hubble Space Telescope offers both ``ultra-disruptive'' and ``disruptive'' target-of-opportunity (ToO) modes. The ultra mode is for $<$2 day turnarounds, but is only allowed once per proposal cycle and cannot be used for far-UV observations (which require more stringent safety checks).  The standard disruptive mode produces response times of $\sim$2--5 days, and can be utilized up to eight times per cycle.  These observations are especially useful for early UV spectroscopic observations where HST is uniquely sensitive and responsive \citep{vasylyev22,Vasylyev26,Bostroem23,Zimmerman24,Wang24}.  Unfortunately, at the current rate of observations, the field is getting  a meager number of early UV spectroscopic observations of nearby SNe: $\sim$1 new CC~SN and no new thermonuclear SN datasets per observation cycle.  Keeping HST operational into the 2030s through a reboost mission will be important to preserve this capability, even with the upcoming UVEX mission, where 8\% of the observing time will be devoted to ToO observations \citep{UVEX}, providing some fast-response UV spectroscopy in the 2030s.

The Swift Observatory Urgency 0 ToO mode, introduced in Guest Investigator Cycle 21, is an exemplar for the needs of the LSST era. These ToOs were responded to autonomously, without a human in the loop, allowing observations to begin within minutes of the request via the Swift API \citep[see ][for technical details]{Swift0,Kennea26}.  Such early UV observations are ideal for shock-breakout observations and for other signs of interaction in early light curves, including both core-collapse and thermonuclear SNe.  There is no currently approved successor to Swift, where multiband UV+optical+Xray observations can be obtained within minutes of discovery, an urgent need for ULTRASAT and UVEX and their high-cadence and/or fast-response observations. %Indeed, the upcoming ULTRASAT mission will identify shock breakouts of nearby core collapse SNe via its high cadence observations of $\sim$200 deg$^2$, potentially 10's per year \citep{ULTRASAT}.  This opportunity will need to be accompanied by significant follow-up to obtain full light curves (and thus trace the early temperature evolution of the explosion).

Also on the horizon is the Schmidt Observatories Lazuli Space Observatory \citep{lazuli}. Lazuli is a planned 3\,m space telescope designed for rapid-response observations ($<$4 hr from trigger), equipped with an $ugriz$ imager and a low-resolution integral field spectrograph (0.4--1.7$\mu$m).  This powerful observatory may execute dozens of rapid follow-up observations per year \citep{Lazuli_tdamm}, potentially collecting large numbers of early SN light curves and spectra capable of extending the distance horizon for young-SN studies.  Lazuli will also be able to obtain spectra of NIR-bright SN precursors \citep{Davies22} out to $\sim$10 Mpc \citep{Lazuli_tdamm}, which would be an important verification of their emission properties.

\subsection{The 80 hour Embargo on LSST Images}

The 80\,hr embargo on full-frame, processed LSST images presents a challenge for the early detection and response to young transients, as it prevents the visual inspection or derivation of forced photometry for undetected transients within this time period.  This will make the search for faint transients around the time of new detections impossible (e.g., to search for low-significance infant emission at the position of a new SN candidate that may be discovered by another survey). We suggest to the National Science Foundation and the Department of Energy that the 80\,hr embargo be waived in regions around nearby galaxies ($\lesssim$100--200 Mpc) to enable the early detection of SNe in nearly real time.

%\subsection{Infrastructure}

%Infrastructure issue.  80 hour

%Finding young, nearby supernovae will require the merging of multiple transient streams beyond LSST to obtain tight non-detections and point to the most promising candidates.  At this point, none of the transient brokers plans to be a `clearinghouse' to merge all the time domain data available into a single data set, even though this would greatly facilitate supernova studies in the nearby universe.  We recommend that one or more brokers take on this challenge.  If necessary, a downstream meta-broker could focus on nearby galaxy science, and only ingest a subset of the full LSST and other transient streams around galaxies within $D$$\lesssim$50-100 Mpc.  Such a focused data broker could also provide other contextual information tailored to the nearby galaxy supernova science case (e.g. space-based archival information for progenitor identification, multi-wavelength catalog data, etc).

\subsection{Disentangling Faint Transients from Young SNe}

\begin{figure}
    \centering
    \includegraphics[width=0.47\textwidth]{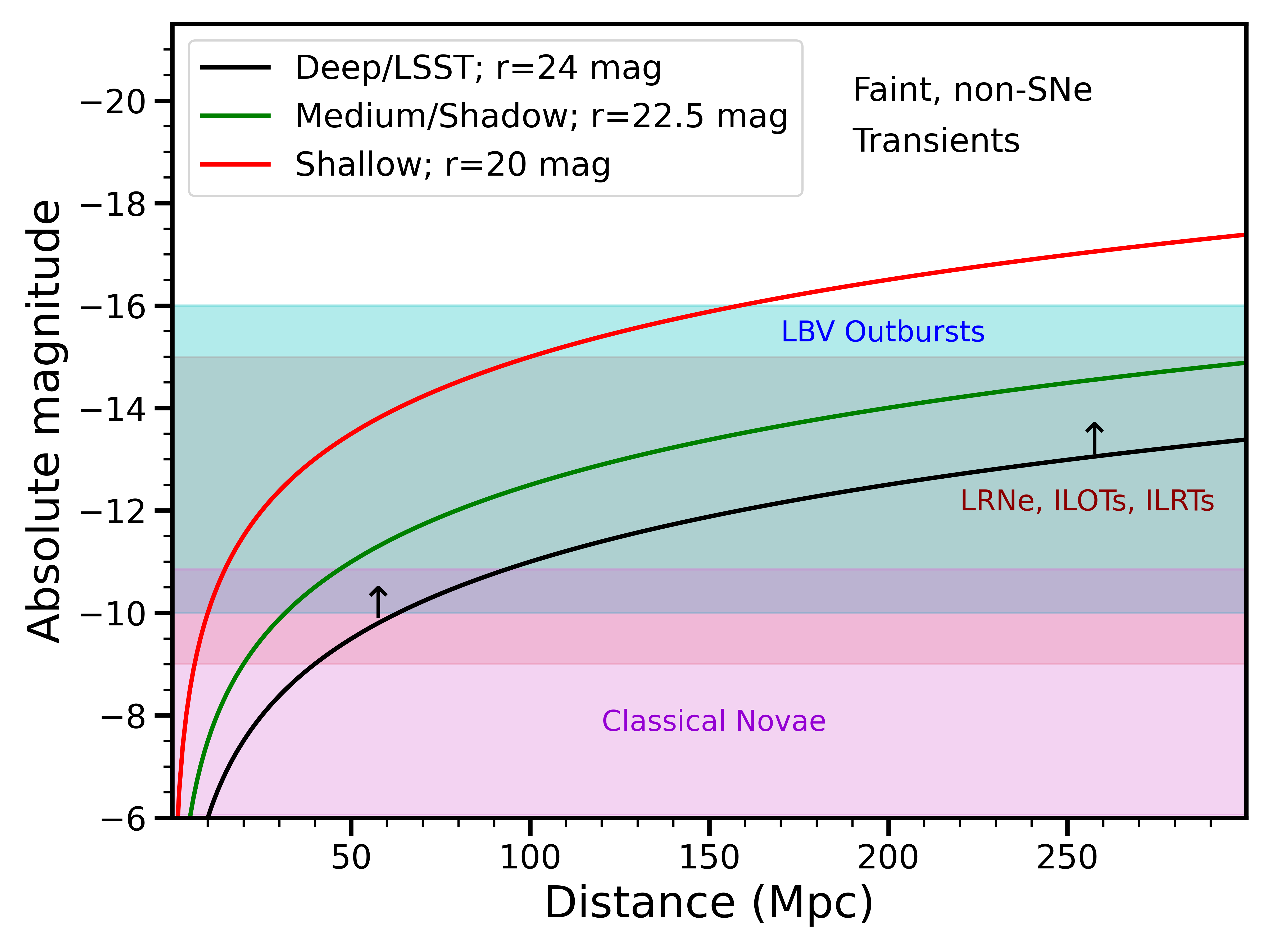}
    \caption{Approximate survey sensitivities for faint, nonterminal transients.  While scientifically important in their own right, the low luminosities will make these transients difficult to disentangle from young, nearby SNe, especially since they will be detectable by the LSST WFD out to large distances.}
    \label{fig:fainttrans}
\end{figure}

Nearly all of the early SN signatures documented in Table~\ref{tab:SNsignatures} and discussed in this work are intrinsically faint, with an absolute magnitude $M_V \gtrsim -16$ mag in the first hours to days after explosion.  This regime overlaps with a growing menagerie of faint transients such as luminous blue variable eruptions, luminous red novae, intermediate-luminosity optical transients, classical novae, and other so-called ``gap transients'' \citep[e.g.,][]{Smith11,Kasliwal12,Pastorello19,Darnley20,Howitt20,TVS23,Ransomeshadow}.  To illustrate, we plot the approximate luminosity range of these transients against LSST and other survey sensitivities in Figure~\ref{fig:fainttrans}.  There is strong overlap, with relatively deep shadowing surveys ($r$$\sim$22.5 depth) capable of detecting bright novae out to $\sim$50 Mpc, while more luminous (but still faint) transients will be detectable out to $\sim$100--200 Mpc (and even farther with LSST itself).

The LSST era will be a boon for these gap transients, illuminating the conditions and physics of stellar eruptions, mergers, and other interactions. While most transient astronomers will be excited to study these faint transients in detail, they will still be confounding for programs focused on young SNe prior to classification.  %Luminous blue variable outbursts have been supernova contaminants for decades, and are sometimes referred to as `supernova impostors' \citep{vandyk00,Smith11,Kochanek12}; other classes of stars may have similar supernova-like outbursts as well \citep[e.g.][]{Andrews21}.
Prompt photometric data alone are insufficient to distinguish between classes, although recent efforts in early light-curve classification are promising \citep[e.g.,][]{rapid,superphotplus,splash}. Early spectroscopic classification will be decisive, but historic variability at the transient site and the larger host-galaxy context may also aid in vetting.

\subsection{A Nearby Galaxy Transient Broker}

Throughout this work, we have pointed out important infrastructure and observing opportunities for deepening our understanding of SN progenitors and explosions in the LSST era, focusing on early-time observations of nearby galaxies.  Many of these suggestions --- merging of transient streams, easy access to archival data and forced photometry across platforms, fast triggering of follow-up observations --- exist in part or as standalone, but should be brought together into a single integrated system dedicated to this science case: a ``Nearby Galaxies Transient Broker.''

There is precedent for such a system. The LSST project has not only selected seven community alert brokers that will ingest the full alert stream, but there are two ``downstream brokers,'' which are focused on a specific science topic and only ingest a portion of the overall stream \citep{Trilling23,Trilling26}.  
Next we sketch the elements of what a nearby galaxy broker might incorporate.

{\it Light-Curve Aggregation -- Transient Stream Merging, Forced Photometry, Archival Light Curves: } Over the next several years, multiple transient streams will be available from LSST, ZTF, Roman, Argus, LS4, ULTRASAT, and potentially other time-domain programs.  Even those programs that do not have a live transient stream regularly report new discoveries and photometry to TNS.  Few (if any) of the LSST transient brokers have the resources to combine the streams from all of these sources, but a Nearby Galaxy Transient Broker would only need to do so for the small regions around the relevant subset of nearby galaxies, combining alert streams from available sources.  In addition, LSST, ZTF, and ATLAS all have some form of ``forced photometry'' service \citep{Shingles21,Graham22,Masci23}, where detection limits in difference images can be pulled via an API and incorporated with light-curve points in real time (the TNS includes nondetections as well), enabling tight constraints on explosion epochs, and real-time searches for precursor emission.

Time-domain data from even earlier epochs (before 2020) are in principle available.  For instance, data from the original Palomar Transient Factory (PTF) and its successor the intermediate PTF are available from the NASA/IPAC Infrared Science Archive (IRSA)\footnote{\url{https://irsa.ipac.caltech.edu/applications/ptf/}}, but require template identification and subtraction for detections of faint SN precursors \citep[e.g.,][]{Reguitti24}.  Data from the Catalina Real-time Transient Survey  \citep{CRTS} is also available via catalog or web query\footnote{\url{http://crts.caltech.edu/}}. Adding such archival time-domain data to a Nearby Galaxy Transient Broker would extend the utility of the service even further, potentially allowing for precursor searches spanning more than a decade.

%With a real-time merged stream set, early light curves can be reconstructed from the full public survey set, and more precise explosion epochs can be determined

{\it Joint Modeling of Multi-survey Light Curves:}  Aggregating survey streams could also be accompanied by tools for joint light-curve modeling that accounts for photometric uncertainties, asynchronous observations, survey-specific passbands, and non-detections. For nearby, young SNe, such tools could help track the rise and color evolution as measurements arrive and prioritize candidates for immediate follow-up observations. Potential starting points include Gaussian process modeling across phase and wavelength \citep[e.g.,][]{pellegrino26}, alongside recent machine-learning methods for probabilistic multiband reconstruction \citep{chaini26}, characterization of irregular light curves \citep{gondhalekar26}, and forecasting \citep{wu2026}, subject to validation for early-time, multi-survey applications. The model predictions should, however, remain clearly distinguishable from observations, and complement rather than replace available light curves.

{\it Further Contextual Information:}  Beyond recent and historical light-curve data, contextual multiwavelength and host-galaxy information is crucial. This may include basic host-galaxy properties, such as robust distance/velocity information, size, and morphology. 
 %If local environmental information is available (i.e. the transient is coincident with a known HII region) that will also be provided.  
 Catalogs of known variable sources can be cross-matched (e.g., AGN, variable star, X-ray, and gamma-ray catalogs). Additional real-time information, such as whether the transient  was in the TESS, Fermi, SVOM, or Einstein Probe field of regard at the time of explosion would also improve situational awareness \citep[e.g. see some of the goals in ][]{across}.

{\it Archival Space-based Imaging:}  Many nearby galaxies have high-resolution imaging from HST and JWST, often over multiple epochs; further imaging is also available from Spitzer.  Any time an SN explodes in a nearby galaxy ($D \lesssim 20$ Mpc, or farther with JWST) with such imaging taken prior to explosion, it presents an opportunity to directly observe the progenitor star system via careful astrometric alignment.  While many of the early observational signals described in this paper shed light on the progenitor system, direct observations are clearly an essential piece in learning about the end states of stars.  In addition to the UV+optical+IR archival observations of the progenitor we describe here, archival X-ray data can also play a role \citep{Nielsen12,Nielsen13,Nielsen14,Kilpatrick18} and should be incorporated into any Nearby Galaxies Transient Broker.

The landscape of direct imaging of progenitor stars is improving, although the samples are still small, and interpretations of the observational data are being debated. The progenitor stars of $\sim$30 hydrogen-rich SNe (SNe~II) have been identified \citep[e.g.,][]{Smartt09, Smartt15RSG,VanDyk12,VanDyk23,Maund13,Kilpatrick_17eaw,Sollerman21,Kilpatrick23,Kilpatrick23ixf,Kilpatrick25,Jencson23,Xiang24} as RSGs of various masses, although the initial-mass range of SN~II progenitors has still not reached consensus, and often estimates disagree even when using the same dataset for the same SN.  The progenitors of other SN subtypes are also coming into view, with a handful of interacting SN\,IIn progenitors \citep[e.g.,][]{Pastorello18,Smith22,Jencson22,Brennan22}, five hydrogen-poor SNe\,IIb with yellow supergiant progenitors \citep[e.g.,][]{vandyk+13,kilpatrick+16}, and two hydrogen-depleted SNe\,Ib with yellow or blue supergiant detections \citep[e.g.,][]{Kilpatrick21}.  Even more exotic, rare SNe now have progenitor candidate imaging as well \cite[e.g., SN Ia-CSM;][]{Szalai26}.

While the space-based imaging mentioned above is generally available in archives, there is additional opportunity for nearby galaxies.  These images should be available on-demand for galaxies where a progenitor could plausibly be identified ($D \lesssim 20$--40 Mpc), already astrometrically aligned, and with point-source catalogs across bands.  With this in-hand, progenitor systems can be identified in realtime and broadcast to the community (with data products) to further inform observational decisions.

%SN Ia-CSM: \citep{Szalai26}

{\it TOM-like Capabilities:} Once all relevant nearby galaxy transient data are available in a single interface, it becomes the natural place to also conduct community-wide follow-up observational campaigns, including the triggering of telescopes, storing of light-curve and spectral data, and even communication between team members.   A community-wide Nearby Galaxy Transient Broker would ideally have data-sharing flexibility in identifying teams and collaborations on the fly. 

A Nearby Galaxy Transient Broker should employ both a web front-end and API on an equal footing, as many users will want to interact with it in an automatic way.  API endpoints to retrieve transient data and to post new data, along with hooks to report back information to the LSST brokers, will ensure overall interoperability with the larger time-domain ecosystem.  Alerts should also be available via an API or Kafka streaming (as a polled or pushed service, respectively), allowing users to poll for new young transients, fast evolving systems, or unusual color evolution, similar to other brokers --- but with the aggregated resources that only a science-focused broker could provide.  

Many of the tools and features described above are partially available through different public platforms but bringing them {\it all} together into a single coherent suite of tools would encourage collaboration and accelerate nearby galaxy transient science.

%Astrometric matching of space-based imaging prior to transient discovery.  Even have catalogs of SN progenitor-candidates on hand.

%TOM-like capabilities of triggering and storing data.

\section{Summary}\label{sec:summary}

This paper has provided a comprehensive overview of the science opportunities that the LSST era will bring for nearby SN science, with an emphasis on early-time observational signatures.  LSST must be combined with other, ongoing transient surveys and experiments in order to obtain higher cadence and earlier detections of young SNe; we summarize the capabilities of current and/or near future transient surveys based on their cadence and depth in Table~\ref{tab:surveys}.  To facilitate LSST shadowing programs, we have categorized vital early SN signatures by their timescale and absolute magnitude in Table~\ref{tab:SNsignatures}, and calculated out to what distance each can be detected in fiducial shadowing programs. A shadow survey of LSST employing deep imaging ($r\approx22.5$ mag; e.g., with DECam or a similarly powerful instrument, \citealt{Ransomeshadow}) could yield hundreds of SNe with early signatures that point to the progenitor system and explosion mechanism if it could cover a significant fraction of galaxies within several hundred Mpc.  We advocate for such an ambitious program. Science highlights from an ideal shadowing survey would include the following.
\begin{itemize}
    \item {\bf Precursor emission from CC~SNe} LSST will dramatically expand searches for precursor outbursts --- novel windows into mass loss and the final stages of stellar evolution --- with potential sensitivity to $\gtrsim$50--100 SNe~IIn and $\gtrsim$5 SNe~Ibn detectable precursors per year.  Given LSST's sensitivity, the fraction of normal SNe~II with SN\,2020tlf-like outburst emission will be known to $\lesssim$10\% within a few years of operations.
    \item {\bf Early spectroscopy of CC~SNe} Rapid identification will enable systematic measurements of short-lived flash-ionization features, constraining the density, composition, and extent of confined CSM.  With $\sim$150 normal SNe~II within $\lesssim$160 Mpc, the incidence of flash-ionization features should be known to $\lesssim$10\% within several years and be connected to the precursor rate.  More intensive spectroscopic campaigns can constrain the duration and evolution of such features, shedding light on the origins of the dense, confined material.
    \item {\bf Early observations of thermonuclear SNe} Within $\lesssim$200 Mpc, there are $\sim$250 Type Ia SNe per year, providing the potential for hundreds of early light-curve excesses over the LSST survey lifetime. Systematic early observations can measure the incidence and diversity of excesses, very high ejecta velocities, and unburned carbon, probing the progenitor environment, ejecta structure, and explosion mechanism. Concentrated campaigns could constrain the timescale and color evolution of early excesses as a function of SN~Ia subtype.
    
\end{itemize}

Realizing this potential will require bold shadowing programs, with robust community-wide follow-up campaigns to gather the requisite datasets. Early SN work may be further facilitated by waiving the 80 hour embargo on LSST images, at least for nearby galaxy fields.  It will also require improvements to the time-domain infrastructure that connects surveys to observers. We therefore advocate for a Nearby Galaxy Transient Broker that brings together transient streams from LSST and complementary surveys, forced photometry and nondetections, long-baseline archival light curves, and contextual information on nearby galaxies. It could also provide rapid access to astrometrically-aligned pre-explosion HST and JWST imaging, enabling progenitor candidates to be identified and distributed to the community in real time. By combining these data products in a science-focused interface with both web and API access, the broker could provide a unified way to identify young and rapidly evolving transients, trigger follow-up observations, and coordinate observations across collaborations.
%To maximize the scientific opportunity, we have also surveyed the infrastructure landscape, and have made concrete 

%In addition to highlighting early supernova science opportunities, we have also surveyed the infrastructure landscape and have identified challenges, both technical (e.g. the LSST 80-hour rule for providing images and forced photometry) and astrophysical (the difficulty of classifying young supernovae and distinguishing faint `gap' transients from terminal explosions). Concrete suggestions for new tools to enable early time science were put forward, many of which could be brought together under the auspices of a `Nearby Galaxy Transient Broker' as an important community resource.  Some aspects may include: 

%\begin{itemize}
%    \item Merging of transient streams
%    \item Easy access to archival time domain data from not only ongoing surveys, but earlier work (e.g. PTF, iPTF, CRTS), to extend pre-cursor searches by decades.
%    \item Easy access to archival high-resolution and other space-based imaging (e.g. HST, JWST, Spitzer) which could identify progenitor system candidates in real time, or even prior to explosion as an outbust is in progress.  This space-based data could be astrometrically aligned to LSST templates prior to any detection, in principle.
%\end{itemize}

The LSST era  offers the opportunity to routinely observe nearby SNe at phases and luminosities that have previously been accessible only rarely and generally by chance. The combination of LSST’s depth, complementary high-cadence shadowing surveys, rapid follow-up monitoring, and integrated time-domain infrastructure can turn the first hours to days of an SN from a largely unexplored phase into a quantitative probe of stellar evolution and explosion physics. Continued development of high-cadence optical surveys and new UV and space-based facilities will further extend this capability, making the coming decade an especially promising time for nearby SN science.

%Several things have not been touched on in this work.  Complementary to the early data we have been emphasizing are late-time data sets.  For core collapse SNe, late-time imagingwhich can be sensitive to late-time CSM interaction  We note that due to the depth of LSST, late time optical light curves will come `for free' for SNe in the nearby universe, and can be stacked in time to obtain extraordinary depths.

%Future facilities: mention ARGUS, ULTRASAT, UVEX.

\begin{acknowledgements}

 Time-domain research by the University of Arizona team and D.J.S. is supported by National Science Foundation (NSF) grants 2308181, 2407566, and 2432036.  This work was performed in part at the Aspen Center for Physics, which is supported by National Science Foundation grant PHY-2210452.
 SN research at Konkoly Observatory is supported by National Research Development and Innovation Office, Hungary, via NKFIH-OTKA grant K142534 as well as the GINOP 2.3.2-15-2016-00003 and GINOP 2.3.2-15-2016-00033 grants from the Hungarian Government, funded by the European Union.
D.H. is supported by STScI grants HST-GO-17770.002, JWST-GO-12468.001, and JWST-GO-09964.001.

Supernova research at Rutgers University is supported by NSF award 2407567 and DOE award DE-SC0010008. S.W.J. also gratefully acknowledges support from a Guggenheim Fellowship.

M.M. and the METAL group at UVA acknowledge support in part from ADAP program grant  80NSSC22K0486, from NSF grant AST-2206657, from HST grant GO-17596 from the Space Telescope Science Institute, which is operated by the Association of Universities for Research in Astronomy, Inc., under NASA contract NAS5-26555. Support was also provided by the NSF under Cooperative Agreement 2421782, and Simons Foundation grant MPS-AI-00010515 awarded to the NSF-Simons AI Institute for Cosmic Origins — CosmicAI, \url{ https://www.cosmicai.org/}.

S.C. acknowledges support received from the NASA FINESST program, grant 80NSSC25K0312.
N.F. acknowledges support from the NSF Graduate Research Fellowship Program under grant DGE-2137419.
S.D. acknowledges support from the NSF Graduate Research Fellowship Program under grant 2444111. Any opinions, findings, and conclusions or recommendations expressed in this material are those of the author(s) and do not necessarily reflect the views of the National Science Foundation.
A.V.F.’s research group at U.C. Berkeley acknowledges financial                 
assistance from Gary and Cynthia Bengier, Clark and Sharon Winslow, Alan Eustace and Kathy Kwan (W.Z. is a Bengier-Winslow-Eustace Specialist in Astronomy), Timothy and Melissa Draper, Briggs and Kathleen Wood, Ellyn and Alan Seelenfreund (T.G.B. is Draper-Wood-Seelenfreund Specialist in Astronomy), and numerous other donors.  
L.A.K. is supported by NASA through an NHFP Hubble Fellowship grant HF2-51579.001-A awarded by STScI.
A.R.L. acknowledges the financial support from Fundação de Amparo à Pesquisa do Estado de São Paulo (FAPESP) through grant 2025/09544-0.
B.H. acknowledges support from NASA through the Arizona/NASA Space Grant Consortium.

\end{acknowledgements}

\vspace{5mm}
%\facilities{Gemini:South (GMOS)}

\software{  astropy \citep{2013A&A...558A..33A,astropy}, SNID-SAGE \citep{snid_sage_2025}
          }

\bibliography{biblio}{}
\bibliographystyle{aasjournal}

\end{document}